\documentclass[twocolumn,aps,prl,preprintnumbers,bibnotes10pt,superscriptaddress]{revtex4}
\UseRawInputEncoding

\usepackage{amsmath}

\usepackage{graphicx}
\usepackage{ulem}
\usepackage{dcolumn}
\usepackage{epsfig}
\usepackage{bm}
\usepackage{array}
\usepackage[frenchb]{babel}
\usepackage{hyperref}
\hypersetup{
colorlinks=true,
citecolor=black,
linkcolor=red,
urlcolor=black
 , bookmarks=true,
 pdfmenubar=true
}

\usepackage{hyperref}
\usepackage{graphicx}
\usepackage{amsmath}
\usepackage{amsfonts}
\usepackage{amssymb}
\usepackage{xcolor}
\usepackage{bbm}
\usepackage{calrsfs}
\usepackage{dutchcal}
\usepackage{amsthm}

\newcommand{\ket}[1]{\ensuremath{|#1\rangle}}

\newcommand{\be}{\begin{equation}}
\newcommand{\ee}{\end{equation}}

\newcommand{\beq}{\begin{eqnarray}}
\newcommand{\eeq}{\end{eqnarray}}

\begin{document}

\title{Frequentist Parameter Estimation with Supervised Learning} 

\author{Samuel P. Nolan}
\affiliation{QSTAR, INO-CNR and LENS, Largo Enrico Fermi 2, 50125 Firenze, Italy}

\author{Luca Pezz$\grave{\text{e}}$}
\affiliation{QSTAR, INO-CNR and LENS, Largo Enrico Fermi 2, 50125 Firenze, Italy}

\author{Augusto Smerzi}
\affiliation{QSTAR, INO-CNR and LENS, Largo Enrico Fermi 2, 50125 Firenze, Italy}

\begin{abstract}

Recently there has been a great deal of interest surrounding the calibration of quantum sensors using machine learning techniques. In this work, we explore the use of regression to infer a machine-learned point estimate of an unknown parameter. Although the analysis is neccessarily frequentist - relying on repeated esitmates to build up statistics - we clarify that this machine-learned estimator converges to the Bayesian maximum a-posterori estimator (subject to some regularity conditions). When the number of training measurements are large, this is identical to the well-known maximum-likelihood estimator (MLE), and using this fact, we argue that the Cram{\'e}r-Rao sensitivity bound applies to the mean-square error cost function and can therefore be used to select optimal model and training parameters. We show that the machine-learned estimator inherits the desirable asymptotic properties of the MLE, up to a limit imposed by the resolution of the training grid. Furthermore, we investigate the role of quantum noise the training process, and show that this noise imposes a fundamental limit on number of grid points. This manuscript paves the way for machine-learning to assist the calibration of quantum sensors, thereby allowing maximum-likelihood inference to play a more prominent role in the design and operation of the next generation of ultra-precise sensors.

\end{abstract}

\maketitle


\section{Introduction}

Quantum parameter estimation is an active area of research, which has both fascinating implications for fundamental science and applications in state-of-the art quantum sensors \cite{PezzeRMP2018, DegenRMP2017}. 
Like many areas of quantum science, quantum parameter estimation can potentially benefit from the application of machine learning techniques \cite{CarleoRMP2019, DunjkoRPP2018, MethaPR2019}, in particular adaptive Bayesian schemes \cite{HentschelPRL2010, HentschelPRL2011, LovettPRL2013, LuminoPRAPP2018, XiaoSR2019, XuNPJ2019, PalittapongarnpimPRA2019, PengPRA2020, SchuffNJP2020, FidererARXIV2020}, improved readout in noisy single-qubit magnetometers \cite{SantagatiPRX2019, AhronSR2019, LiuJPHYSB2019, QianAppPhysLett2021, KhanahmadiPRA2021} and state preparation \cite{HainePRL2020, SchaferMLST2020}. 
Although current efforts to improve the sensitivity and utility of quantum sensors is focused on the use of non-classical states \cite{PezzeRMP2018}, the development of sophisticated data analysis techniques to extract information encoded in complex quantum states is also an important aspect of this effort. 
High quality device calibration is essential for such efforts~\cite{LanePRA1993, PezzePRL2007, OlivaresJPB2009, KrischekPRL2011, XiangNATPHOT2011, PezzeBOOK2014, LiENTROPY2018, RubioJPC2018, CiminiPRA2020}.
In particular, the calibration of a generic quantum sensors through the lens of supervised machine learning offers interesting possibilities \cite{GreplovaARXIV2017, CiminiPRL2019, CiminiARXIV2020, NolanARXIV2020}.

Supervised learning is any kind of machine learning technique that makes use of pre-labelled training data. 
The objective is to use these data to infer a labelling rule (or model) $\Theta_W(\boldsymbol{\mu})$ that maps an input $\boldsymbol{\mu}$ (known as feature vector-- in practice given by experimental data) to the desired output $\theta$. Throughout this manuscript, we use the notation that $W$ is a set of model parameters, to be adjusted during the training process. 
We argue that the calibration of parameter estimation experiments is itself a supervised learning task: typically one collects sequences of calibration measurements $\boldsymbol{\mu}$ at many (known) parameter values $\theta_j$ (called the labels), with $j$ running over a set of grid-points. 
This data-set is then used (e.g. via a trained neural network) to infer a function that accepts an arbitrary set of measurement results not used during calibration, and returns an estimate of the (unknown) parameter $\theta$ that has produced them. 
In the language of parameter estimation the labelling rule $\Theta_W(\boldsymbol{\mu})$ is called an estimator \cite{KayBOOK, LehmannBOOK, PezzeBOOK2014}. 


\begin{figure}[t!]
\centering
\includegraphics[width=\columnwidth]{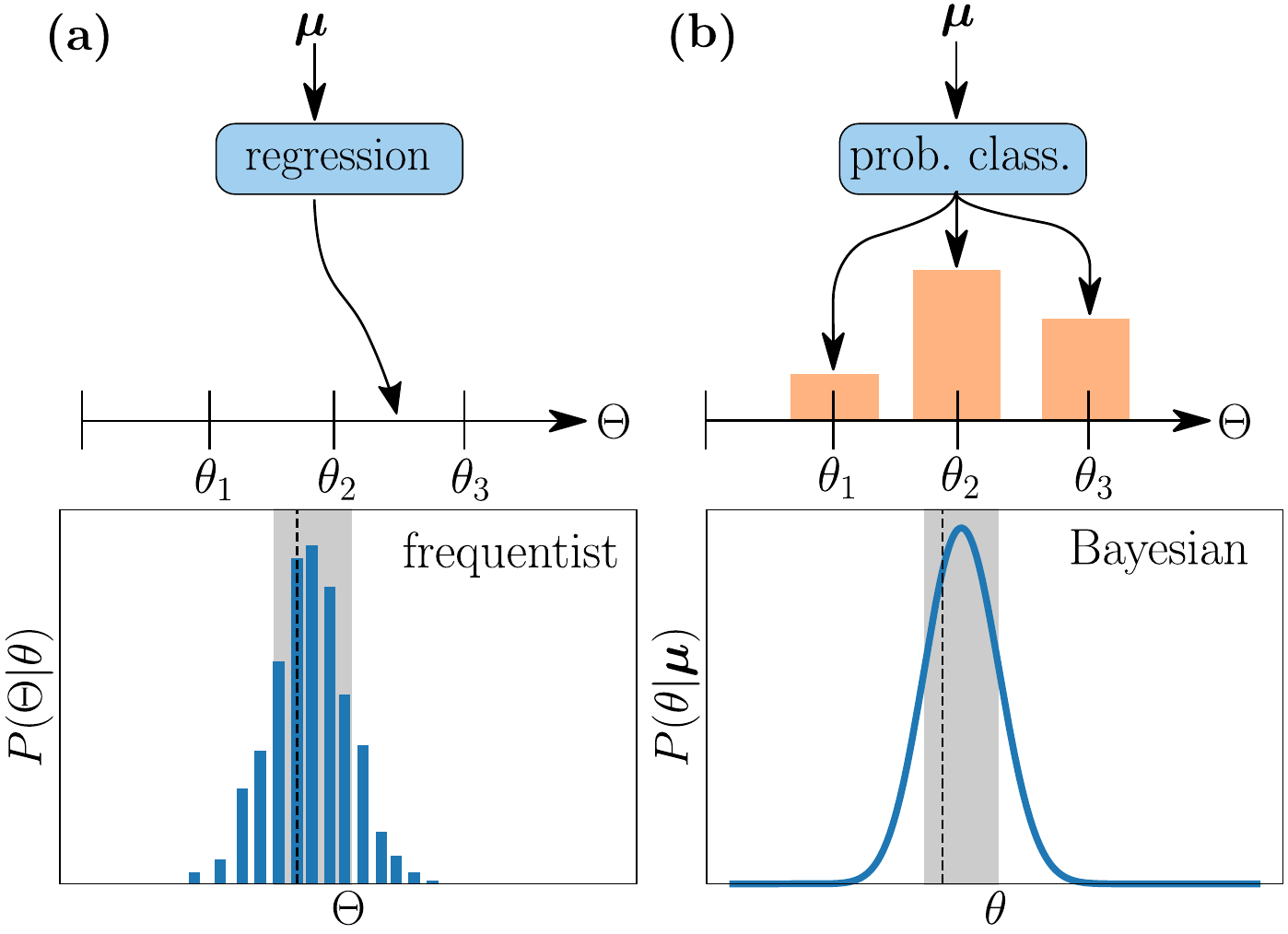}
\caption{\textbf{Schematic of supervised learning in relation to parameter estimation.} The goal of supervised learning to map a feature vector $\boldsymbol{\mu}$ onto a label $\theta$ by training a model $\Theta_W(\boldsymbol{\mu})$. These labels could be either smooth (regression, \textbf{(a)}) or discrete (classification, \textbf{(b)}). As no information about the underlying probability distribution is available, regression necessitates frequentist inference. A histogram is obtained from many estimates $\Theta$ (bottom left). The uncertainty is the statistical fluctuations in $\Theta$ (shaded gray), and for an unbiased estimator the mean converges to the true value (black dashed line). Another approach is probabilistic classification (right), whereby the model provides a conditional probability that a particular category is correct, given $\boldsymbol{\mu}$. These coarse-grained probabilities can be used to perform Bayesian inference (bottom right). The black-dashed line is the true value, and shaded gray region is a Bayesian confidence interval.
}
\label{fig:scheme}
\end{figure}

Broadly speaking, there are two kinds of supervised learning tasks: {\it classification and regression} \cite{CarleoRMP2019, DunjkoRPP2018, MethaPR2019}. 
Regression aims to make predictions on smooth or continuous quantities, and is the focus of this manuscript. 
In regression problems the labels can take on continuous values, and the labelling rule $\Theta_W(\boldsymbol{\mu})$ is a smooth curve that should provide accurate predictions on $\theta$ that don't necessarily coincide with one of the grid values.
As the model can only return a point estimate of the parameter, this approach necessitates a frequentest analysis, whereby repeated estimates must be performed and analysed to obtain statistics. 
The mean of the resultant histogram are taken as the final estimate, and its width the uncertainty - the result of inevitable statistical fluctuations (Fig.~\ref{fig:scheme}, \textbf{(a)}). 
Although not the focus of this work, we provide a brief discussion of classification for context. 
Classification is well suited to naturally discrete or coarse grained data - such as the classification of handwritten images \cite{NielsenBOOK}.
The calibration of quantum sensors has also been formulated as a classification problem, whereby a grid of parameter values $\theta_j$ are treated as categories \cite{GreplovaARXIV2017, NolanARXIV2020} (Fig.~\ref{fig:scheme}, \textbf{(b)}). 
In Ref. \cite{NolanARXIV2020}, the machine-learned confidence intervals assigned to each $\theta_j$ where shown to approximate the Bayesian posterior distribution $P_W(\theta_j \vert \boldsymbol{\mu})$ present in the training set.
Given $P_W(\theta_j \vert \boldsymbol{\mu})$, one can extract a Bayesian estimate $\Theta_W(\boldsymbol{\mu})$ (for instance given by the maximum or the mean value of the posterior distribution) and assign a parameter uncertainty to a confidence interval around this value [e.g. the shaded region in Fig.~\ref{fig:scheme} \textbf{(b)}].
In this sense, classification naturally performs Bayesian inference, and regression naturally performs frequentist inference \cite{note_class}.

In this manuscript we explore regression as a means for calibration of quantum sensors.
We address some interesting and important elements of a regression-based calibration scheme that were not analyzed in previous studies \cite{GreplovaARXIV2017, CiminiPRL2019, CiminiARXIV2020}: we clarify the properties of the estimator $\Theta(\boldsymbol{\mu})$ and its relation to the distribution of the training data. It was previously shown that a subjective prior distribution over the label set plays a key role in Bayesian estimation based on classification \cite{NolanARXIV2020}). Is the same true in a frequentest scheme based on regression?
In particular, we use concrete examples to demonstrate that the optimal estimator is typically the Bayesian maximum a-posteriori estimator, when the number of training measurements is relatively  small. When the number of training measurements is large, we demonstrate that the machine-learned estimator converges to the maximum-likelihood estimator. 
We thus explore conditions under which the machine-learned estimator is unbiased and saturates the ultimate Cram{\'e}r-Rao sensitivity bound. Furthermore, we argue that this bound applies also to the cost function used during training, and if it can be computed a-priori, could be used to aid in the selection of optimal model and training parameters.

Aside from focusing on the rather pragmatic problem of calibrating ultra-sensitive devices, our work also raise fascinating questions about the nature of supervised learning itself. 
Indeed, device calibration typically involves fairly small feature spaces relative to the size of the training set. It thus operates in a regime very different from the usual applications of machine learning, such as image recognition, for example, where
the size of the feature space is 
extremely large in comparison to the number of training examples: in which case the likelihood function $P(\boldsymbol{\mu} \vert \theta)$ is unknowable for all practical purposes, making maximum-likelihood or Bayesian inference impossible. 

To assist the reader, here we provide a table of commonly used notation. 

\begin{center}
\begin{table}
\begin{tabular}{ c|c} 
  &  \\ 
  \hline
$ \boldsymbol{\mu}$ & sequence of independent measurements  \\ 
 $\mu$ & an individual measurement result  \\ 
 m & no. measurements in each $\boldsymbol{\mu}$ for training   \\ 
  $\nu$ & no. measurements in each $\boldsymbol{\mu}$ for estimation  \\ 
$\theta_j$ & element of the training grid (labels)   \\
$M_j$ & no. feature vectors $\boldsymbol{\mu}$ sampled at $\theta_j$ for training \\ 
 $d$ & no. uniformly spaced grid points $\theta_j$  \\ 
 $L$ & size of grid ($\theta_0=0$)  \\ 
 MLE & maximum-likelihood estimator  \\ 
 MAP & Bayesian a-posteriori estimator  \\ 
 \end{tabular}
\caption{Summary of frequently used notation.}
\label{notation}
\end{table}
\end{center}

\section{Background: Regression and Parameter Estimation}

When a ``labelling rule'' $\Theta_W(\boldsymbol{\mu})$ (or model) is trained, its free parameters $\{ W \}$ are adjusted, for example using stochastic gradient descent \cite{NielsenBOOK, MethaPR2019}, the aim of which is to minimise a cost or loss function. 
A common choice is the mean-square error,
\begin{equation} \label{eq:cost}
C=\overline{ \left( \Theta_W(\boldsymbol{\mu}) -\theta \right)^2} 
\end{equation}
where the bar denotes an average over the whole training set. The question central to this manuscript is the following: when performing regression, which labelling rule (or estimator) $\Theta_W(\boldsymbol{\mu})$ would we expect the machine to learn, and why? In this section we provide a brief discussion of frequentist inference, and then provide a simple argument intended to highlight why a particular labelling rule may be learned over another, based on the distribution of the training data.

{\it Frequentist inference.} The goal of parameter estimation is to infer a value of the parameter $\theta$ given a set of observations $\boldsymbol{\mu}$. A central concept is that of an estimator $\Theta(\boldsymbol{\mu})$ - a monotonic function which maps a $\boldsymbol{\mu}$ onto the parameter space. We wish to clarify that the subject of this manuscript is frequentist inference, although this does not preclude the use of Bayesian estimators, which can depend on any available prior knowledge $P(\theta)$ available about $\theta$ in the absence of any information $\boldsymbol{\mu}$. 
However, because any regression model learns the Bayesian point estimate $\Theta_W$ directly, we are forced to rely on frequentist inference strategies.

The frequentist paradigm relies on repeated observations, that are used to gain information about the distribution $P(\Theta \vert \theta)$ (as in Fig.~\ref{fig:scheme}, \textbf{(a)}). The estimate is taken as the mean over all possible measurement sequences of fixed length $m$ 
\begin{equation}
\theta_{\rm est} =  \sum_{\boldsymbol{\mu}} \Theta(\boldsymbol{\mu}) P(\boldsymbol{\mu} \vert \theta) \equiv \langle \Theta(\boldsymbol{\mu}) \rangle (\theta) .
\end{equation}
In practice, this average would typically be approximated using a re-sampling technique such as bootstrapping. It is useful to define the bias of an estimator $\langle \Theta(\boldsymbol{\mu}) -\theta \rangle$. An estimator is said to be unbiased if its bias vanishes $\langle \Theta(\boldsymbol{\mu}) \rangle = \theta$ and $\partial_\theta \langle \Theta(\boldsymbol{\mu}) \rangle = 1$ for all $\theta$, however in general there is no guarantee that an estimator is unbiased. 
The uncertainty in the estimate is merely the result of inevitable statistical fluctuations, e.g. a common choice of risk function is the variance
\begin{equation} \label{eq:freqvar}
\Delta^2\theta(\theta)= \langle \left( \Theta(\boldsymbol{\mu})-\langle \Theta \rangle \right)^2 \rangle.
\end{equation}
The variance of any unbiased estimator is bounded asymptotically by the Cram{\'e}r-Rao bound (CRB) $\Delta^2\theta \geq \Delta^2\theta_{\rm CRB} = 1/{m F}$, where $F$ is the Fisher information 
\begin{equation} \label{eq:Fisher}
F(\Theta) = \left. \sum_\mu \frac{\left[\partial_\theta P(\mu \vert \theta) \right]^2}{P(\mu \vert \theta)} \right|_{\theta = \Theta}.
\end{equation}

\begin{figure}[t!]
\centering
\includegraphics[width=\columnwidth]{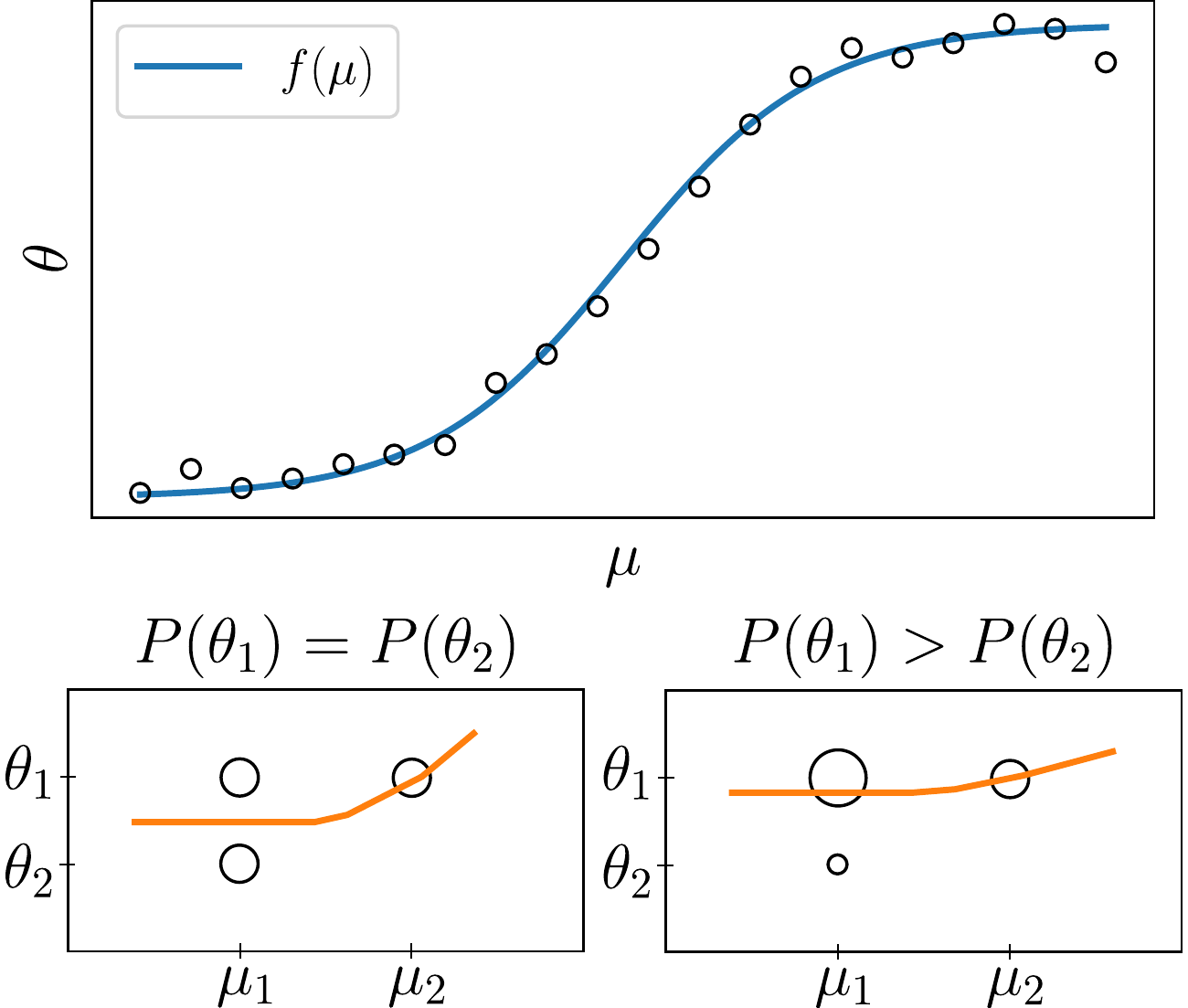}
\caption{\textbf{Simple illustration of regression.} \textbf{(top)} For a one-dimensional input $\mu$, we illustrate a set of training examples sampled from a Gaussian likelihood function Eq.~(\ref{eq:gauss_likelihood}). The training set (black circles) lie along the curve $f(\mu)$ (blue line) which maximises the likelihood function, with some scatter $\sigma$. Naively, given enough points one would expect a fit to reproduce this curve. \textbf{(bottom)} The prior distribution $P(\theta)$ can also influence the fit (orange line), if for example the same $\mu$ is observed at multiple $\theta$. Here, the size of the circle is proportional to the number of observations.}
\label{fig:regression}
\end{figure}

{\it Model training.} Now we return to the question posed at the start of this section: which estimator might we expect a model learn? Consider a pair of random variables $(\boldsymbol{\mu},\theta)$, with some unknown relationship defined by the conditional probability $P(\boldsymbol{\mu} \vert \theta)$ (called the likelihood function. In Fig.~\ref{fig:regression} we provide a simple illustration of a set of points $(\mu_j, \theta_j)$ sampled from $P(\mu \vert \theta)$, for the special case that the feature vectors $\boldsymbol{\mu}$ are one dimensional. These points may lie along a curve $f(\boldsymbol{\mu})$, which must be encoded somehow in $P(\boldsymbol{\mu} \vert \theta)$. The goal of a regression algorithm is to use the training points to infer a model $\Theta_W(\boldsymbol{\mu})$ that does a good job of reproducing $f(\boldsymbol{\mu})$. 
As an example, the likelihood function may be modelled as a Gaussian
\begin{equation} \label{eq:gauss_likelihood}
    P(\boldsymbol{\mu} \vert \theta) \propto \exp \left(-\frac{\left[ \theta - f(\boldsymbol{\mu}) \right]^2}{2 \sigma^2} \right).
\end{equation}
Such a likelihood function would produce training points that lie along the curve defined by $f(\boldsymbol{\mu})$ with some scatter given by $\sigma$, as illustrated in Fig.~\ref{fig:regression} (top panel).
At least for this example, one would expect any reasonable fit to learn to reproduce the maximum $\Theta_W(\boldsymbol{\mu}) \approx f(\boldsymbol{\mu})$. In a parameter estimation problem, the function $f(\boldsymbol{\mu})$ should be monotonic.

However, the likelihood function is not the whole story: we argue that the prior knowledge $P(\theta)$ over the label set is also relevant. 
Suppose the labels are first sampled from $P(\theta)$. 
If the same $\boldsymbol{\mu}$ value is sampled at multiple $\theta$ values, (Fig.~\ref{fig:regression}, bottom panels), then the relative frequency of each observation may also influence any possible fit to the training data. This argument is based only on the distribution of the training data, but should hold for any reasonable training and model choices. 
Specifically we assume (1) a sufficiently large training set (2) the model $\Theta_W$ has enough flexibility to learn the ``target'' function $f$ with reasonable fidelity and (3) the training algorithm has sensible hyperparameters. Becaue the model $\Theta_W$ can depend on a prior distribution $P(\theta)$ over the label set, in this sense we argue that regression is naturally Bayesian. 


In summary, any machine-learning algorithm is trained on input-output pairs sampled from the joint distribution \cite{DunjkoRPP2018, MethaPR2019}
\begin{equation} \label{eq:joint}
P(\boldsymbol{\mu}, \theta ) = P(\boldsymbol{\mu} \vert \theta ) P(\theta) ,
\end{equation}
and it is the properties of this distribution that determine the optimal  $\Theta_W(\boldsymbol{\mu})$.
Although $P(\theta)$ is a probability distribution over the label set, this does not necessarily imply that $\theta$ is a random variable: within the Bayesian interpretation of probability $P(\theta)$ is a statement of knowledge or certainty, and is chosen subjectively by the user to include any knowledge about $\theta$, prior to the collection of any evidence $\boldsymbol{\mu}$. 

Throughout this manuscript, we assume that the feature vectors $\boldsymbol{\mu} = \{\mu_1, \mu_2, \cdots \mu_m \}$ contain $m$ independent elements, which are denoted without boldface.
In many supervised learning applications, these elements are not independent (think of an image converted into an array of pixel data; the ordering is crucial). 
We find it convenient to instead work with the relative frequencies
\begin{equation} \label{eq:freq}
f_\mu = m_\mu/m
\end{equation}
of individual measurement outcomes $\mu$. 

{\it Familiar estimators.}
A common Bayesian estimator is the value of the parameter $\theta$ that maximises the Bayesian posterior distribution for a given $\boldsymbol{\mu}$, called the maximum a-posteriori (MAP) estimator 
\begin{align} \label{eq:MAP1}
& \Theta_{\rm MAP}(\boldsymbol{\mu}) = \mathrm{argmax}_\theta \left[P(\theta \vert \boldsymbol{\mu}) \right] \\
&= \mathrm{argmax}_\theta \left[ \sum_\mu f_\mu \log [ P(\mu \vert \theta) ] + \frac{\log[ P(\theta)]}{m}\right] \label{eq:MAP2} 
\end{align}
where the sum runs over each unique measurement outcome $\mu$, not each possible sequence $\boldsymbol{\mu}$. 
Here, 
\begin{equation} \label{eq:Bayes}
P( \theta \vert \boldsymbol{\mu})  = \frac{\prod_{j=1}^m P(\mu_j \vert \theta) P(\theta)}{P(\boldsymbol{\mu})}
\end{equation}
is the Bayesian posterior which follows from the symmetry of the joint distribution $P(\boldsymbol{\mu}, \theta) = P(\theta, \boldsymbol{\mu})$ \cite{LehmannBOOK, PezzeBOOK2014}. We have used the fact that the likelihood function $P(\boldsymbol{\mu} \vert \theta)$ factorises for $m$ independent measurements.
The conditional probability $P( \theta \vert \boldsymbol{\mu}) $ is the posterior distribution, and can be thought of as an update of the prior knowledge, given new evidence $\mu$. 
The denominator $P(\boldsymbol{\mu})$ is fixed by normalisation.

When working exclusively in the frequentist formalism, the notion of prior knowledge is not meaningful, and thus another common estimator is the maximum-likelihood estimator (MLE)
\begin{align} \label{eq:MLE1}
& \Theta_{\rm MLE}(\boldsymbol{\mu}) = \mathrm{argmax}_\theta \left[P( \boldsymbol{\mu}  \vert \theta) \right] \\
&= \mathrm{argmax}_\theta \left[ \sum_\mu f_\mu \log [ P(\mu \vert \theta) ] \right] . \label{eq:MLE2} 
\end{align}
Notice that $\Theta_{\rm MAP}$ and $\Theta_{\rm MLE}$ coincide for flat priors $P(\theta)$, and for sufficiently well-behaved priors when $m \rightarrow \infty$. Within the Bayesian formalism, the MLE can be understood as the asymptotic limit of the MAP estimator. 
Indeed, in the limit that $m \rightarrow \infty$, some fairly general analytic results are available regarding the Bayesian posterior distribution. Using the fact that for independent measurement results the likelihood function factorises,  
it is possible to show that the Bayesian posterior becomes a Gaussian \cite{LehmannBOOK, PezzeBOOK2014} 
\begin{equation} \label{eq:post_asymptotic}
P(\theta \vert \boldsymbol{\mu}) = \sqrt{\frac{m F(\Theta)}{2\pi}} \exp \left(-m F(\Theta)[\theta - \Theta(\boldsymbol{\mu})]^2/2 \right).
\end{equation}
The maximum $\Theta$ of the Bayesian posterior is the MAP estimator Eq.~(\ref{eq:MAP2}), which converged to the MLE as $m \rightarrow \infty$. 

Of course, neither the MLE or MAP estimators can be computed unless the likelihood function $P(\boldsymbol{\mu} \vert \theta)$ is known. For this reason, it is always necessary to calibrate the estimation experiment, which typically requires a large set of training measurements. This is precisely an act of supervised learning, which brings us back to the question posed at the start of this section. If an estimator $\Theta_W(\boldsymbol{\mu})$ is trained (or calibrated) based on a large set of training points sampled from the joint distribution Eq.~(\ref{eq:joint}), which estimator would we expect to learn, and why? When the number of training measurements are large, following the argument in Fig.~\ref{fig:regression} the Gaussian form of Eq.~(\ref{eq:post_asymptotic}) would imply that the MAP should be the expected estimator. When the number of training measurements are small, no such general argument exists. However, in the following section we shall demonstrate that the answer is still very often the MAP estimator.

\section{Numerical Results}

\begin{figure}[t!]
\centering
\includegraphics[width=0.8\columnwidth]{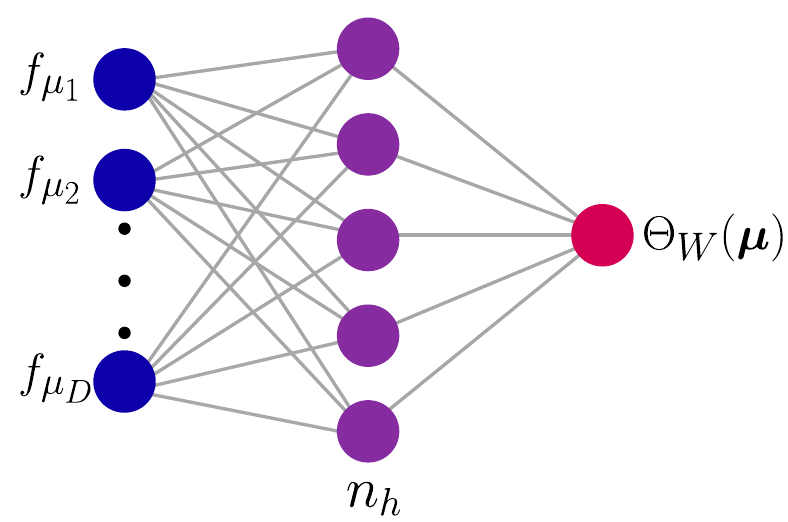}
\caption{\textbf{Artificial neural network architecture.} Throughout this manuscript we use this architecture to train a model able to map measurement results converted to relative frequencies $\boldsymbol{\mu} = \{ f_1, f_2, \cdots f_D \}$ to a phase estimate $\Theta$. The number of hidden neurons $n_h$ is related to the complexity of the model, introducing more free parameters $W$. The output neuron is able to take on any continuous value, hence this realizes a regression model.}
\label{fig:NN}
\end{figure}

{\it Artificial neural network.} In this section we explore frequentist parameter estimation using artificial neural networks (ANNs). 
A trained ANN is a function $\Theta_W(\boldsymbol{\mu})$ of measurement data $\boldsymbol{\mu}$, depending on many free parameters $\{ W \}$ that are adjusted during training and characterize the ANN. The power of ANN models lies in their universality: ANNs can approximate any function with arbitrary accuracy, limited only by the size of the network \cite{NielsenBOOK, MethaPR2019}. As such, they are well suited to problems where the form of $\Theta_W(\boldsymbol{\mu}) $ is unknown. We do not discuss the details of ANNs here, for pedagogical overview the reader is referred to Ref. \cite{NielsenBOOK}. The network design is depicted in Fig.~\ref{fig:NN}. For a $D$-dimensional feature space (namely, $D$ indicates the number of possible measurement outcomes), measurement results are tallied at a fixed (but unknown) $\theta$. The resulting array of relative frequencies $\boldsymbol{\mu} = \{ f_1, f_2, \cdots f_D \}$ is fed into the model. It is important to use relative frequencies (and not the raw tallies) because the ANN is better able to generalise to any number of measurements. The output is a point-estimate $\Theta_W(\boldsymbol{\mu})$ of the parameter $\theta$.


For the training set we define a uniform parameter grid consisting of $d$ points over the relevant domain $\theta_j \in \left[0,L \right]$, where $j=1, \cdots d$. The grid spacing is $\delta \theta = L/(d-1)$
The training measurements are performed as follows. First, a label $\theta_j$ is sampled randomly from the prior $P(\theta_j)$. A feature vector (measurement sequence) $\boldsymbol{\mu}$ of size $m$ is then sampled at this $\theta_j$ from the likelihood function $P(\boldsymbol{\mu} \vert \theta_j) = \prod_{k=1}^{m} P(\mu_k \vert \theta_j)$. In our numerical study  $P(\boldsymbol{\mu} \vert \theta_j)$ is known to us but never seen by the model. This process is repeated until $M_{j}$ feature vectors  $\left \{ \boldsymbol{\mu}_{1}, \boldsymbol{\mu}_{2} \cdots \boldsymbol{\mu}_{M_j} \right\}$ are sampled at each label $\theta_j$. If $m$ is sufficiently large, a re-sampling technique such as bootstrapping could be used (instead of considering several $M_j$ feature vectors) if the total number of measurements is an important resource~\cite{CiminiPRL2019}. However, we do not employ such a bootstrapping strategy here. Once the training data has been collected, a random set of initial weights $W$ are chosen for the model. For reasons of efficiency, it is common to divide the data into randomly chosen mini-batches, and compute the cost Eq.~(\ref{eq:cost}) only over the mini-batch (rather than the whole training set). Using the cost the parameters $W$ are updated e.g. using gradient descent \cite{NielsenBOOK}. Once the entire training set has been used to compute an update of the model parameters $W$, an epoch is said to have passed. Typically a model is trained for many epochs. Throughout this manuscript we use the python-based package Keras \cite{chollet2015keras} to build and train our ANN models. All networks have $D$ input neurons (for a $D$-dimensional Hilbert space), $n_h$ hidden neurons with the ``ReLU'' activation function, and a single output with linear activation, as in Fig.~(\ref{fig:NN}). Training is done using the ADAM algorithm for stochastic gradient descent, which aims to minimise the mean-square error cost Eq.~(\ref{eq:cost}).

{\it Single qubit benchmark example.}
The likelihood function fully characterises the sensor, and is central to parameter estimation. Below, as a pedagogical example, we consider the case of a single qubit prepared in the state $\ket{\uparrow}$.
The qubit is rotated by an angle $\theta$ in the Bloch sphere, such that $|\psi_\theta \rangle = \exp(-i \hat{\sigma}_y \theta/2)| \uparrow \rangle$ is the state after the rotation.
Finally we project over the $\sigma_z$ basis resulting in two possible outcomes:
$\uparrow$ or $\downarrow$,
with $\theta$-dependent probabilities $P(\uparrow | \theta) = \cos(\theta/2)^2$ and $P(\downarrow | \theta)=1-P(\uparrow | \theta)$, respectively.
These probabilities are analogous to a classical biased coin with (generally) unbalanced head and tail events.
For this likelihood function, the CRB and MLE can be computed analytically. The CRB is $\Delta^2 \theta = 1/m$ for all $\theta$, which is also known as the standard quantum limit (SQL). The MLE is
\begin{equation} \label{eq:qubitMLE}
\Theta_{\rm MLE}(f_\uparrow) = 2\arccos \left( \sqrt{f_\uparrow}\right) .
\end{equation}
This estimator is monotonic on the phase interval $\theta \in \left[ 0, \pi \right]$. This system is particularly valuable for building intuition due to its simplicity. In particular, it is one-dimensional - as revealed by Eq.~\ref{eq:qubitMLE} the estimation problem is determined by a single number, the relative frequencies of a single measurement result, here ``$\uparrow$''. 
Aside from being a purely pedagogical example, NV-centre magnetometry is a well-known application of a single-qubit quantum sensor \cite{DegenRMP2017}. However, in these devices calibration is not the key challenge - typically the likelihood function can be modelled explicitly \cite{BonatoNATNANO2015}.

In the following, we first present numerical results for this example, obtained in the non-asymptotic regime of a relatively small number of training data $m$. We then discuss the opposite limit or $m \gg 1$, which the relevant regime for calibrating quantum sensors.


\begin{figure}[t!]
\centering
\includegraphics[width=\columnwidth]{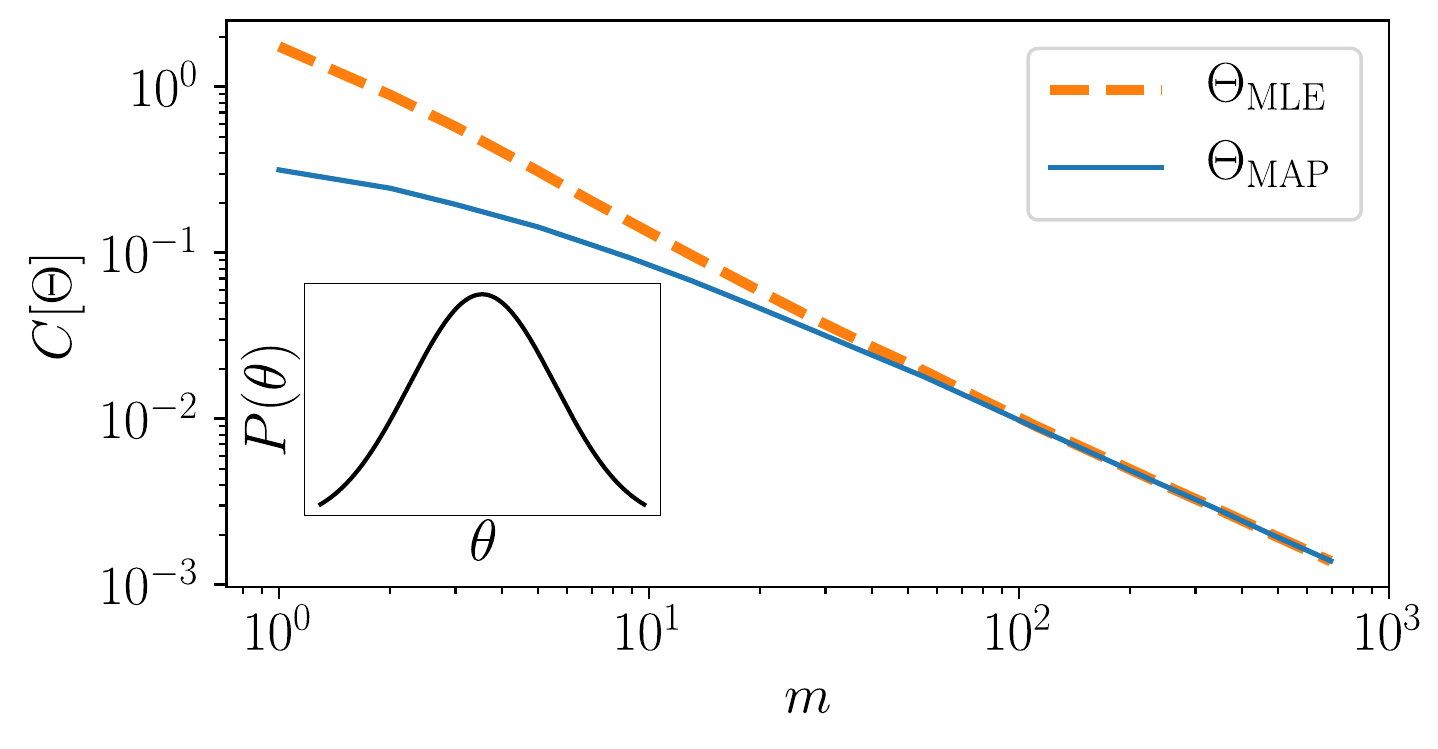}
\caption{\textbf{Impact of the prior on the cost function.}
Mean-square error cost function Eq.~(\ref{eq:cost}) as a function of the number of training data $m$ for the single-qubit example with a non-flat prior (explicitly shown in the inset). Here we manually compute the cost function for both the MLE (dashed orange line) and MAP (solid blue line) estimators on a training set with $d=10$ grid-points and $\overline{M_j}=5000$ feature vectors at each (on average). When $m$ is small, the cost function will favour maximising the Bayesian posterior distribution, rather than the likelihood function. Interestingly, the cost function favours the Bayesian MAP estimator over the MLE even though the training data contains no information about the total number of measurements $m$ - only the frequencies are used. As expected, when $m$ becomes large the two estimators become indistinguishable to the cost function. }
\label{fig:bayesian_costs}
\end{figure}


\begin{figure*}[t!]
\centering
\includegraphics[width=\textwidth]{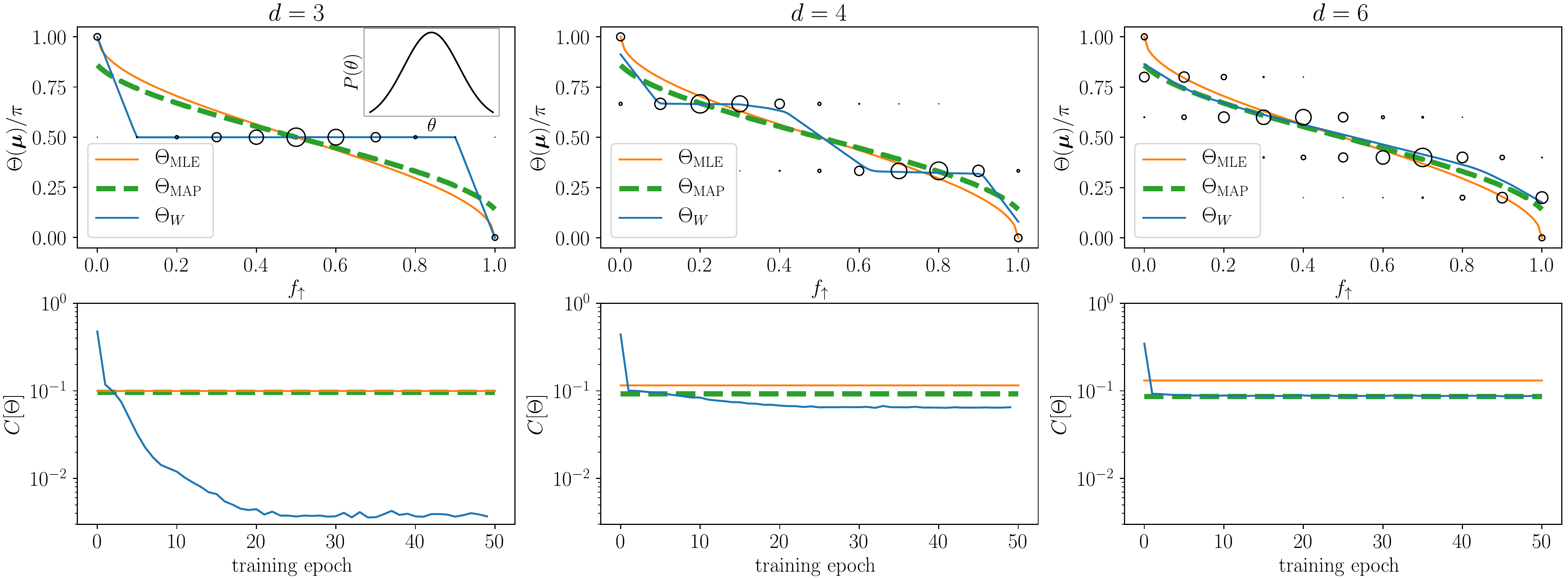}
\caption{\textbf{When $m$ is small, ANNs can learn the MAP estimator, but when $d$ is small they tend to over-fit.} \textbf{(top row)} We compare the machine-learned estimator $\Theta_W$ (solid blue), to the Bayesian MAP estimator $\Theta_{\rm MAP}$ (computed numerically) and the MLE $\Theta_{\rm MLE}$ Eq.~(\ref{eq:qubitMLE}). Data was sampled from a Gaussian prior \textbf{(inset)} shown with $m=10$, on a training grid with $d$ points in $\theta \in [0,\pi]$. We visualise the training data as hollow circles, whose relative sizes are proportional to the number of times a particular measurement result $f_{\uparrow}$ occurred at that $\theta_j$ relative to the most common result. When $d$ is small, the ANN tends to overfit, learning an estimator that performs well on the training set but will obviously generalise poorly (such as the $d=3$ results here). As $d$ increases, the model appears to learn the a good approximation to the MAP estimator. An indication that overfitting has occurred is that the model's cost function is far below its value when evaluated for MLE and MAP estimators (horizontal lines). The ANN models use a single input neuron taking the value $f_{\uparrow}$ and one hidden layer with $n_h=1024$ neurons.  The training set contains $\overline{M_j}=10^3$ samples on average with $m=10$ at each grid-point. During training a mini-batch size of 32 was used.}
\label{fig:bayesian_est}
\end{figure*}

\subsection{Non-asymptotic results}

First we train networks in the regime where each feature vector $\boldsymbol{\mu} = \{ f_{\uparrow}, 1-f_{\uparrow} \}$ contains only a small number of measurements $m$ for each $\theta_j$. The number of feature vectors $M_j$ sampled at each $\theta_j$ may still be large. In principle, the networks only needs a single real number $f_{\uparrow}$ as input - however we have found that better results are obtained using two inputs: $ f_{\uparrow}$ and $1-f_{\uparrow}$.  When $m$ is small, the effect of a non-flat prior knowledge $P(\theta_j)$ over the label set can be significant, notice that the MLE and MAP estimator differ Eq.~(\ref{eq:MLE2}) only by a prior-dependent term that scales like $1/m$. To illustrate this difference, in Fig.~\ref{fig:bayesian_costs} we manually evaluate the cost function over the training set sampled from a non-flat prior (i.e. no networks are trained here). This confirms that the mean-square error cost Eq.~(\ref{eq:cost}) does indeed favour the MAP estimator. Interestingly, the training set contains only frequencies, which lack any explicit information about $m$, however the MAP estimator cannot be computed without knowledge of $m$ [see Eq.~(\ref{eq:MAP2})] Despite this, Fig.~\ref{fig:bayesian_costs} demonstrates that the cost function favours the MAP estimator over the MLE when the prior $P(\theta)$ is non-flat. As expected the difference in cost between the MAP and MLE estimators vanishes as $m \rightarrow \infty$, confirming that to understand the role of $P(\theta_j)$, it is necessary to study the small $m$ limit. Of course, the case of small $m$ is not relevant from an estimation point of view, since the estimator is generally strongly biased in this regime.

In Fig.~\ref{fig:bayesian_est} we show the machine-learned estimator $\Theta_W$ as a function of $f_{\uparrow}$ compared to the MLE and MAP estimators for a Gaussian prior $P(\theta) \propto \exp[-(\theta-\pi/2)^2 ]$ (Fig.~\ref{fig:bayesian_est}, inset).
We train the ANN model using three different training sets with $d=3,4,6$ grid-points using the broad Gaussian prior at $m=10$.

The hollow circles are a visualisation of the training set. The diameter of each circle is proportional to the number of times that $f_{\uparrow}$ was observed at a particular label $\theta_j$ relative to the most frequent outcome in the entire training set - in other words, they are proportional to $P(\theta \vert f_{\uparrow})$. Along the bottom row we plot the cost during training, compared to the cost computed for the MLE and MAP estimators manually over the training set (as in Fig.~\ref{fig:bayesian_costs}). 

With $d=3$ grid points, the ANN learns a highly biased estimator that performs well over the training set, but generalises poorly to unseen values of $\theta$. This is indicated by a cost that is much lower than the expected cost, achieved by learning an estimator that is  mostly flat. The point is that although Fig.~\ref{fig:bayesian_costs} indicates that the MAP estimator is favoured over the MLE, at least for this training set the highly-biased estimator depicted in Fig.~\ref{fig:bayesian_est} ($d=3$) has a far lower cost than either. In other words, this model performs well on the training set, but generalises poorly. In the language of machine learning, this is called overfitting \cite{MethaPR2019}.

Increasing the number of grid-points, combats overfitting to some extent. For example, even including a single additional grid-point $d=4$ brings the model much closer to the MAP estimator (as indicated by comparing the costs). This is caused by a kind of competition between neighbouring points - due to intrinsic noise $\Delta \theta$ in the training data, increasing $d$ while holding $m$ fixed inevitably leads to the same input $f_{\uparrow}$ being observed at different $\theta_j$. 
This spread $\Delta \theta$ in the training data is visible in the training set in Fig.~\ref{fig:bayesian_est} (hollow circles), notice that both $\theta=0,\pi$ represent eigenstates of the measurement observable and therefore have $\Delta \theta=0$. When neighbouring points ``overlap'' to a large degree, in order to minimise the cost the model is forced to interpolate between these points, and this is precisely when the prior has a significant effect (see Fig.~\ref{fig:regression}). As a result, a mostly flat estimator, such as $d=3$, is no longer optimal. Finally, with only $d=6$ grid-points the model appears to learn a good approximation to the MAP estimator, which is clearly distinct from the MLE at $m=10$. 
As discussed in Fig.~\ref{fig:bayesian_costs}, notice that the cost function evaluated for the MAP estimator is smaller than that of the MLE due to the non-flat prior.


\begin{figure}[t!]
\centering
\includegraphics[width=\columnwidth]{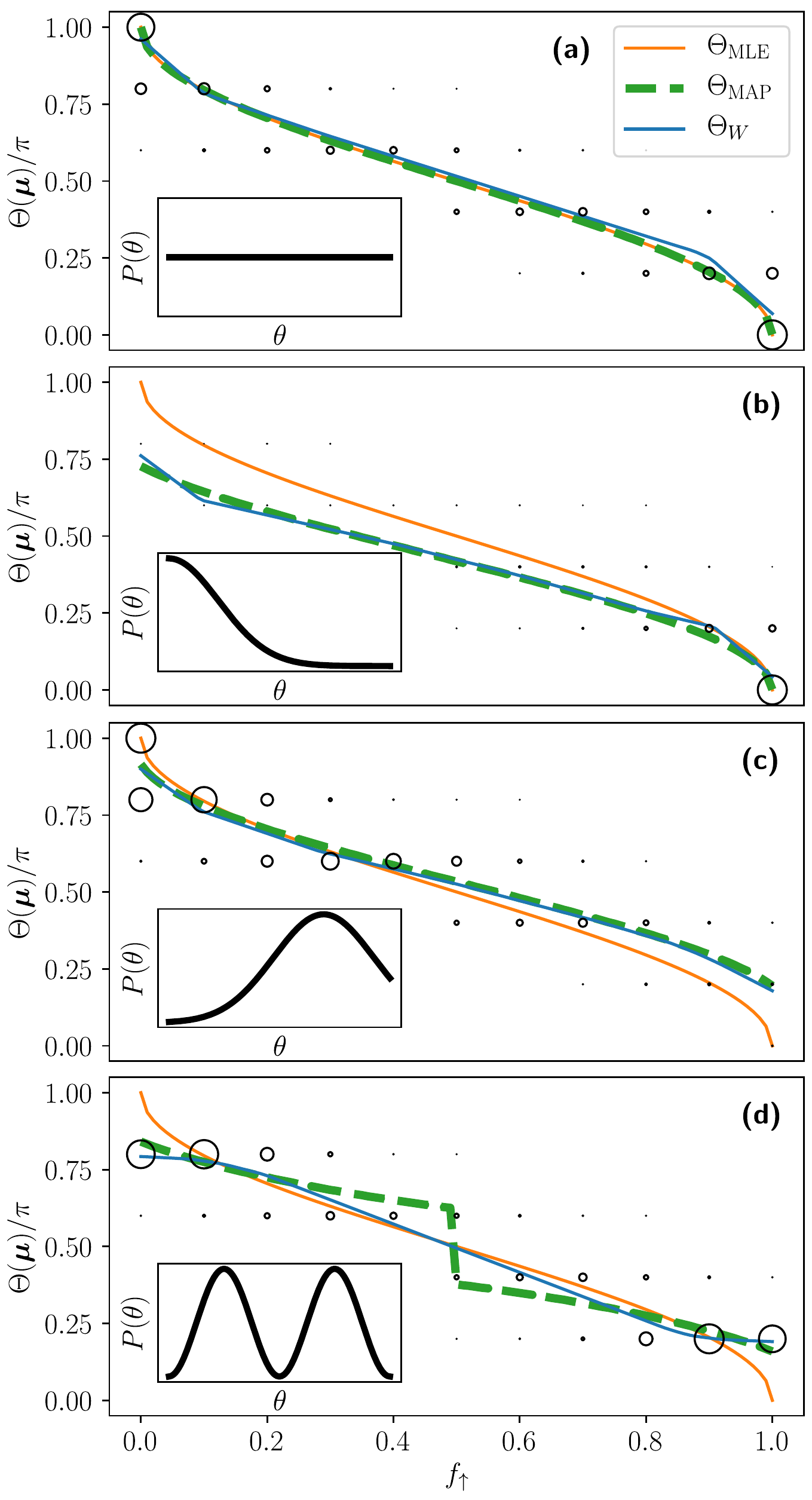}
\caption{\textbf{Comparing models with different priors.} We demonstrate that the machine-learned estimator $\Theta_W$ learns a good approximation to the Bayesian MAP estimator for some other priors \textbf{(a-c)}, and also provide a counter-example \textbf{(d)}. All models were trained with $d=6$ grid points and $m=10$, and otherwise the same training set and model architecture as Fig.~\ref{fig:bayesian_est}. The points are a visualisation of the training set, the relative diameters are proportional the to the number of observations. Notice that when an observation $f_{\uparrow}$ appears at multiple phases, the relative number of observations (proportional to the prior) determines fit.}
\label{fig:compare_priors}
\end{figure}

Even aside from overfitting, the MAP estimator is not always optimal. In fact, the model only appears to learn an approximation to the MAP estimator 
\be
\Theta_W \approx \Theta_{\rm MAP}
\ee
when the Bayesian posterior distribution $P(\theta_j \vert f_{\uparrow})$ is sufficiently ``regular'', namely smooth and single-peaked.  In Fig.~\ref{fig:compare_priors} we explore the machine-learned estimator $\Theta_W$ for several priors, again for $d=6$ grid points with $m=10$. With $m=10$, the priors chosen in Fig.~\ref{fig:compare_priors} (a-c) yield single-peaked $P(\theta_j \vert f_{\uparrow})$, and we confirm that in these cases, the machine-learned estimator learns a good approximation to the maximum Bayesian estimator $\Theta_W \approx \Theta_{\rm MAP}$. Fig.~\ref{fig:compare_priors} (d) provides a counter-example: the machine-learned estimator does not converge to the MAP estimator. In this case, the simple picture presented in Fig.~\ref{fig:regression} does not hold - there is not a well-defined curve $f(f_{\uparrow})$ present in the distribution of training points.

\subsection{Asymptotic Results}

In this section we turn our attention to the training of ANN models in the asymptotic limit $m \rightarrow \infty$, and also study the usefulness of such models for performing frequentist inference after they have been trained. To avoid confusion, we clarify here that the models are trained on a sequences of $m$ measurements during training, and after training, the model is used to perform inference on a sequence of $\nu$ measurements at a fixed but unknown phase $\theta$, which is to be estimated. Crucially, $m$ and $\nu$ need not be equal, and in fact any useful model should generalise well to $\nu$ not seen during training.

\begin{figure}[t!]
\centering
\includegraphics[width=\columnwidth]{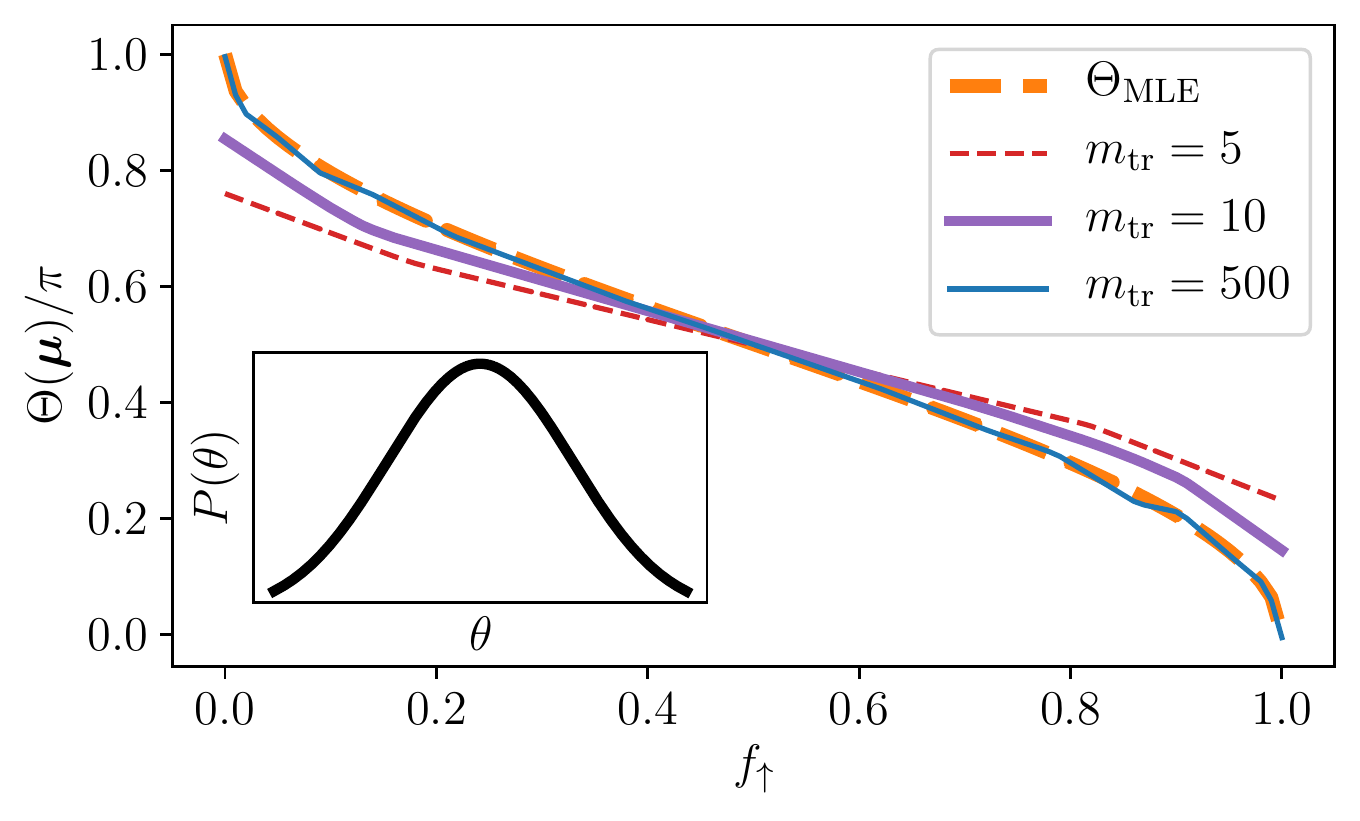}
\caption{\textbf{Asymptotically in $m$ the machine-learned estimator converges to the MLE.} For a single qubit with a Gaussian prior (inset), we train three models with $m=5,10$ and 500. As $m$ increases, $\Theta_W$ appears to converge to the MLE. Models were trained on $d=10$ grid-points with $\overline{M_j}=10^3$ feature vectors each. Gradient descent was run for 50 training epochs with a mini-batch size of 32. Models have a single hidden layer with $n_h=1024$ neurons.}
\label{fig:estmtr}
\end{figure}

In the previous section we found that so long as the Bayesian posterior distribution has a single, well-resolved peak, an ANN estimator will learn a good approximation to the MAP estimator. When the number of training measurements is large, the Bayesian posterior converges to a Gaussian distribution and the MAP estimator converges to the MLE. Thus, we can conclude that when $m \rightarrow \infty$, under ideal training conditions an ANN estimator should learn a good approximation to the MLE. This result is demonstrated in Fig.~\ref{fig:estmtr} for the single-qubit example introduced in the previous section. For a Gaussian prior, we plot the machine-learned estimator for increasing $m$ and observe that it converges to the MLE. More precisely, in Fig.~\ref{fig:compare_priors}(a-c) we found that for a Gaussian posterior distribution that is not over-fit, the machine-learned estimator is a close to the MAP estimator $\Theta_W \approx \Theta_{\rm MAP}(m)$, which depends on the number of training measurements $m$. As $m \rightarrow \infty$ we have 
$\Theta_{\rm MAP} \rightarrow \Theta_{\rm MLE}$. This implies

\be
\Theta_W \to \Theta_{\rm MLE},
\ee
and is precisely what is observed in Fig.~\ref{fig:estmtr}.
In particular, in a ``well-trained'' network (see discussion below) the parameter sensitivity achieved by the estimator should saturate the CRB.


\subsection{The asymptotic role of prior knowledge.}
 
Here we investigate the impact of the prior distribution in the case of a large number of measurements.
Crucially, the asymptotic result Eq.~(\ref{eq:post_asymptotic}) is {\it prior-independent}, but this is not to say that the prior knowledge can never play a role asymptotically. For instance, Eq.~(\ref{eq:post_asymptotic}) only holds if the prior is non-zero and non-singular around $\Theta_{\rm MLE}$, which need not always be the case. 

\begin{figure}[t!]
\centering
\includegraphics[width=\columnwidth]{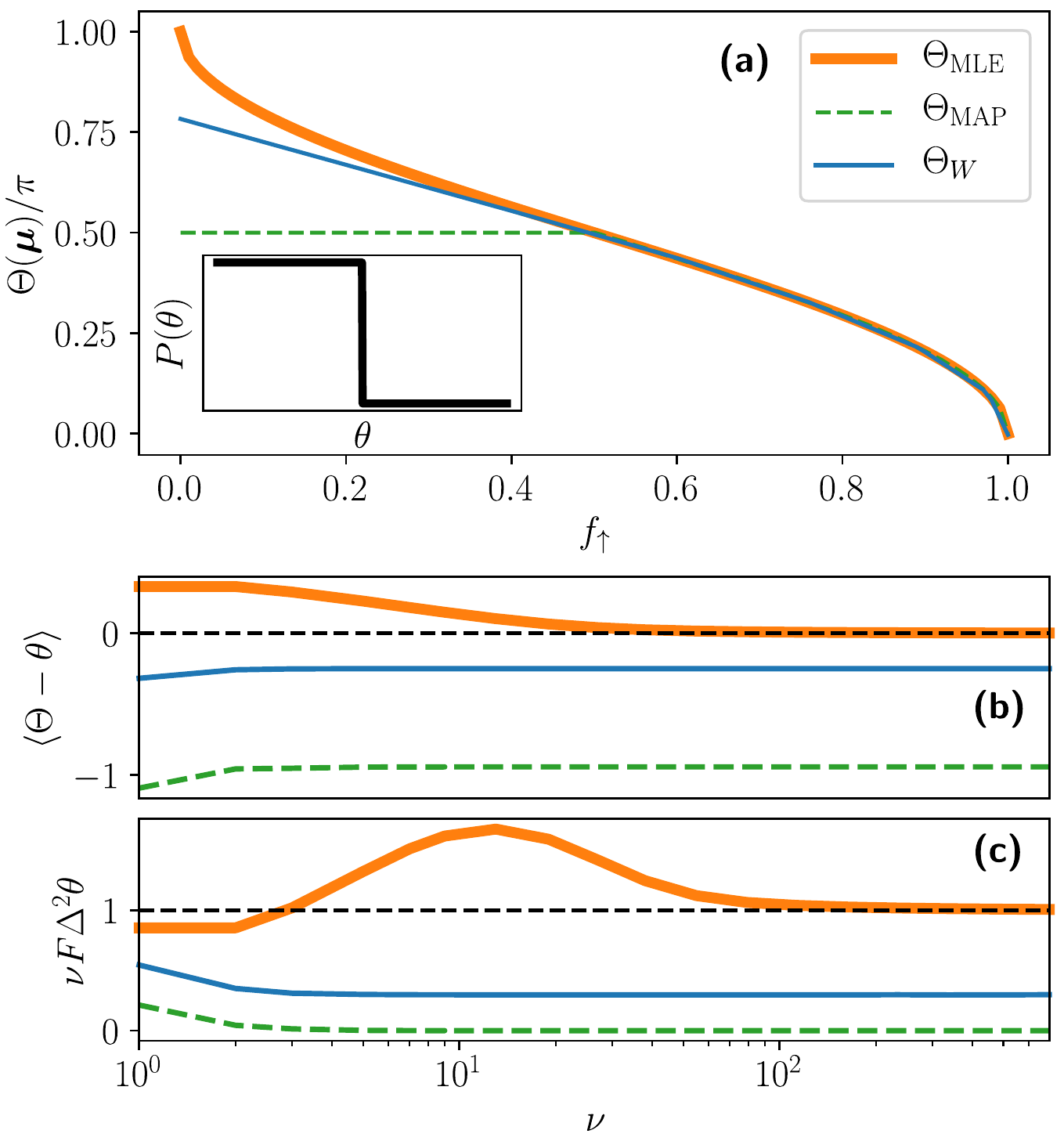}
\caption{\textbf{Step-function prior.} For a single qubit we train a model using a step-function prior, resulting in a training set containing no examples from half the grid $\theta \in [\pi/2, \pi]$. Unlike Fig.~\ref{fig:compare_priors} the training set is firmly in the asymptotic limit with $m=10^4$ \textbf{(a)} The machine-learned estimator $\Theta_W$ is compared to the MLE and MAP estimators, and in \textbf{(b,c)} we plot the bias and variance, normalised to the CRB as a function of $m$ at $\theta = 0.8 \pi$, deliberately chosen to test the model's performance in the grid-half it was never shown. The result is a highly biased estimator, even as $m \rightarrow \infty$. We use a single hidden layer with $n_h=1024$ neurons. The training set has $d=5$ grid-points in $\theta \in \left[0, \pi/2 \right]$, with $\overline{M_j}=10^3$ feature vectors at each grid point, on average. During training a mini-batch size of 8 was used for a total of 50 trainig epochs. }
\label{fig:step}
\end{figure}

First, we consider a step-function prior. During training, the model is never shown data from half the grid $\theta \in [\pi/2, \pi]$ (Fig.~\ref{fig:step} (a), inset). The resultant model $\Theta_W$ is a good approximation to the MLE over the grid-half that it was shown ($\theta \in [0, \pi/2]$) but for $\theta > \pi/2$ the model attempts to generalise, Fig.~\ref{fig:step} (a), but the result is a highly biased estimator that is neither the MLE or the MAP. After training, if the model is used to estimate a phase $\theta \in [\pi/2, \pi]$, the resultant estimate is biased Fig.~\ref{fig:step} (b) but with reduced variance Fig.~\ref{fig:step} (c), even as $\nu \rightarrow \infty$. Compare this to the MLE, which is asymptotically unbiased and saturates the CRB.


\begin{figure*}[t!]
\centering
\includegraphics[width=\textwidth]{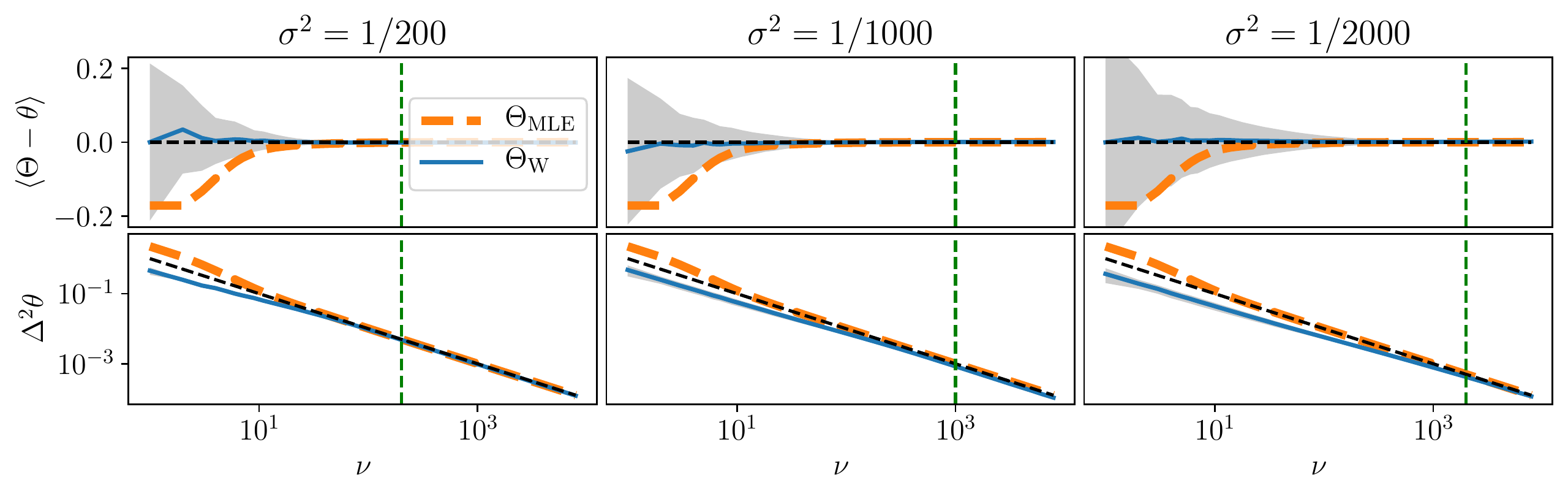}
\caption{\textbf{Saturation of the CRB with non-flat priors.} We plot the mean bias (top row) and variance (bottom row) of an estimate at $\theta = 0.4 \pi$ as a function of $\nu$, with a single qubit. The goal is to demonstrate that the machine-learned estimator has the correct asymptotic properties (unbiased and saturates the CRB) for narrow Gaussian priors with variance $\sigma^2$ and mean $\theta_0=0.4 \pi$. We train 10 models with a single hidden layer of $n_h=1024$ neurons, and show the mean (solid blue line) and standard deviation (shaded gray regions) compared to the MLE (orange dashed line). Vertical dashed green lines indicate the prior variance $1/\sigma^2$. The training set contained $d=50$ uniformly spaced grid-points between $\theta \in \left[0, \pi \right]$ with Gaussian prior $P(\theta) \propto \exp \left[-(\theta - \theta_0)^2/2\sigma^2 \right]$. The mean number of feature vectors used at each phase was $\overline{M_j}=10^3$ with $m = 10^4$ measurements at each. During training we use a mini-batch size of 128 for 10 training epochs.}
\label{fig:prior_asymptotic}
\end{figure*}

Although step-function priors may be a little contrived, we include the example to clearly make the following point: even asymptotically, prior knowledge plays an inescapable role in the supervised training of an estimator $\Theta_W$. To be more quantitative, we now consider a narrow Gaussian prior, $P(\theta) \propto \exp \left( -\left[\theta - \theta_0 \right]^2/2 \sigma^2 \right)$. If the prior is sufficiently narrow, it will produce similar results to the step function prior because there will be many $\theta$-values rarely (or never) shown to the model during training. In Fig.~\ref{fig:prior_asymptotic} we verify that even when this is the case, the machine-learned estimator has the correct asymptotic properties (unbiased and saturates the CRB as $\nu\rightarrow \infty$). We train several models with different $\sigma$, and use them to estimate $\theta_0$. The variance and bias are plotted as a function of $\nu$, revealing that so long as $\nu \sim 1/\sigma^2$ or greater, the estimate is roughly unbiased and saturates the CRB. When $\nu \ll 1/\sigma^2$ the estimates are inconsistent - large fluctuations are observed in the trained models, which have high bias and reduced variance. Note that in Fig.~\ref{fig:prior_asymptotic} we use $m=10^4$ for which the MLE and MAP estimators are extremely close - however due to the narrow prior the model is never shown some $\theta$-values, and similarly to the step function prior Fig.~\ref{fig:step}, learns a distinct estimator. As a general rule we might say that for well-behaved priors, the Fisher information of the prior
\begin{equation} \label{eq:conditionprior}
    \nu \gtrapprox \frac{1}{F(\theta)}\int d\theta' \frac{[\partial_{\theta'} P(\theta')]^2}{P(\theta')}
\end{equation}
provides a threshold for the optimal use of the machine-learned estimator.

\section{Optimal Training Parameters}

The discussion presented thus far has been concerned with the properties of the machine-learned estimator $\Theta_W$, based on the statistics of the training set when the number of training data $m$ are large. 
From the Gaussian form of the Bayesian posterior distribution Eq.~(\ref{eq:post_asymptotic}) in the limit where $m \rightarrow \infty$, we argue that one would generally expect to learn the MLE, as demonstrated in Fig.~\ref{fig:estmtr}. 
In Fig.~\ref{fig:costsCRB} we show that when $m \rightarrow \infty$, the convergence of the cost to the phase-averaged CRB can be an indication that the model is well-trained. After the model is trained, the goal is of course to use it to estimate an unknown value of the parameter $\theta$ based on a sequence of $m$ independent measurements. Crucially, $\nu$ and $m$ need not be the same. In this section, we discuss some conditions for optimal performance of the machine-learned model $\Theta_W$. Specifically, up to some $\nu$ it should be (1) unbiased, within the inevitable statistical fluctuations $\Delta^2 \theta$ and (2) these fluctuations should be close to the CRB.

No model can truly be expected to perform well for arbitrary $\nu$ - at a certain point one must expect the model to break down. Here we show that the density of the training grid is the key factor that appears to limit the model's ability to generalise to large $\nu$. However, minimum useful training grid density is itself limited by quantum noise in the training data.

\subsection{When is the network ``Well-trained''?}

\begin{figure}[t!]
\centering
\includegraphics[width=\columnwidth]{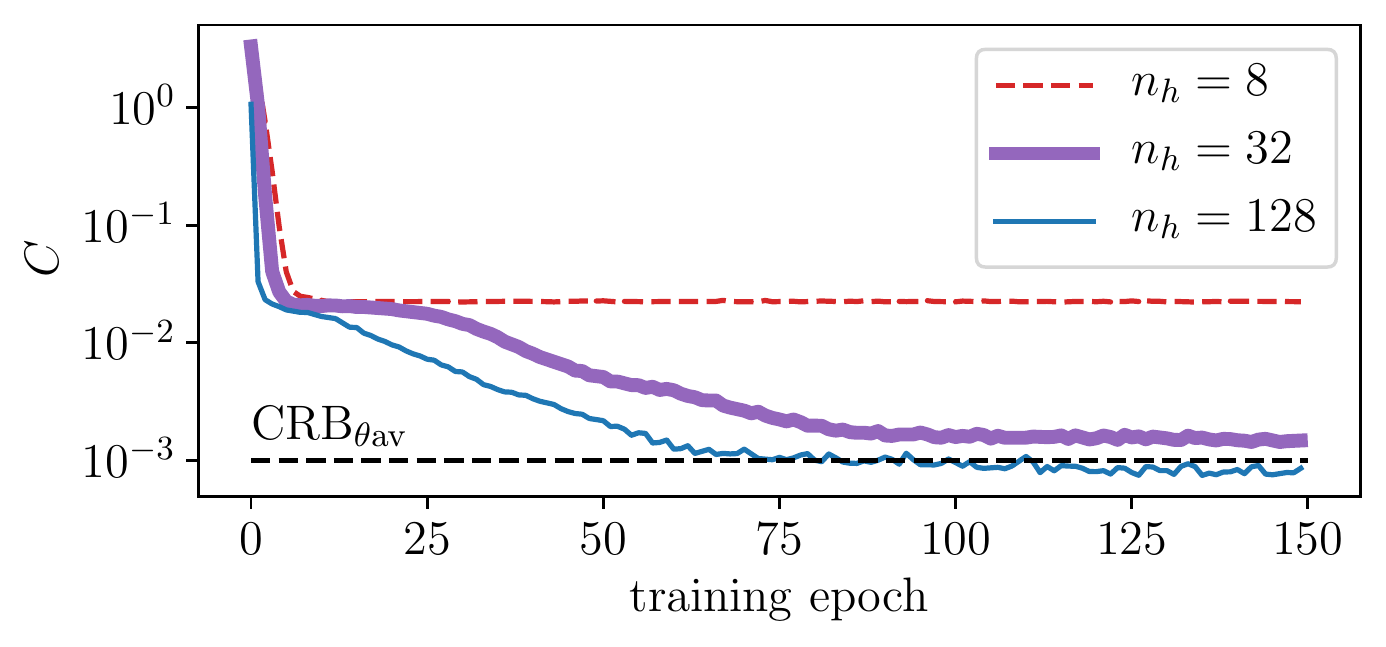}
\caption{\textbf{Cost function converges to the phase-averaged CRB.} For a single qubit in the asymptotic regime ($m = 10^3$), we train three different ANN models with the same training set. As the number of hidden neurons $n_h$ increases (all model have only a single hidden layer), the cost-function approaches the phase-averaged CRB as the training set is iterated over. An epoch is a single training iteration over the entire training set. We use a mini-batch size of 8 on a training set with $d=10$ uniformly spaced grid-points between $\theta \in \left[0, \pi \right]$ sampled from a flat prior. The mean number of feature vectors used at each phase was $\overline{M_j}=10^2$.}
\label{fig:costsCRB}
\end{figure}

Of course, ideal training conditions can be difficult to achieve in practice. However, it is possible to exploit the fact that asymptotically in the number of training measurements the MLE aught to be the optimal model to help choose good training hyper parameters. The key observation is that the mean-square error cost function, Eq.~(\ref{eq:cost}) should also be bounded by the CRB. If the number of feature vectors $M_j$ sampled at a particular $\theta_j$ is large, the sample average used to compute the cost should converge to the expected value $C(\theta_j) = \overline{\left(\theta_j - \Theta_W(\boldsymbol{\mu})  \right)^2} \rightarrow \langle (\theta_j - \Theta_W(\boldsymbol{\mu})  )^2 \rangle$ as $M_j \rightarrow \infty$. Here the overline denotes an average over all samples $\boldsymbol{\mu}$ collected at a single $\theta_j$. This quantity is the mean-square error for the estimator $\Theta_W$, which is also bounded by the Cram{\'e}r-Rao bound \cite{MSE}. We introduce the notation 
\begin{equation} \label{eq:CRBphaseav}
\mathrm{CRB}_{\theta \mathrm{av}} = \sum_{j=1}^d \Delta^2 \theta_{\rm CRB}(\theta_j) P(\theta_j)
\end{equation}
for the phase-averaged CRB. If one expects that the optimal model should be close to the MLE, then the cost should saturate the phase-averaged CRB. Unlike the mean-square error (as in Fig.~\ref{fig:bayesian_est}, which provides a tighter bound), the phase-averaged CRB has the advantage of being estimator independent. If the CRB can be computed, the phase-averaged CRB can be used to ensure that a particular choice of model and training algorithm is optimal given the training data, so long as (1) $m$ is sufficiently large that the asymptotic result Eq.~(\ref{eq:post_asymptotic}) applies and (2) the training set is sufficiently large that sample averages approximate expectation values (i.e. large $M_j$). Of course, the phase-averaged CRB is not a strict inequality - aside from the obvious possibility of overfitting (as in Fig.~\ref{fig:bayesian_est}), the CRB is an asymptotic bound that can be violated from below for finite $m$. Furthermore, the CRB is a bound on the expectation value, however the cost is computed as a sample mean, which is prone to additional statistical fluctuations. Despite this, if the CRB can be computed a-priori (for instance, any separable quantum system is bounded by the standard quantum limit (SQL) $\Delta^2 \theta_{\rm CRB} \geq \Delta^2\theta_{\rm SQL} = 1/{m} N$, where $N$ is the number of probes), the phase-averaged CRB can be a useful heuristic for designing models and choosing training hyperparameters. For instance, in Fig.~\ref{fig:costsCRB} we demonstrate that phase-averaged CRB can be used to select the number of neurons in the network's hidden layer $n_h$, and the number of training epochs. Roughly speaking, increasing $n_h$ increases the number of free-parameters $\{ W \}$ to be trained, but also provides the model with additional flexibility. In Fig.~\ref{fig:costsCRB} if $n_h$ is too small, the cost plateaus early during training, indicating that perhaps the model cannot learn the MLE as well as possible, given the training set. However, saturation of the phase-averaged CRB does not imply that a given model will necessarily generalise well to unseen inputs - a training set that contains only a single $\theta$ point will always generalise poorly.


\subsection{The role of grid resolution relative to $\nu$.}

\begin{figure}[t!]
\centering
\includegraphics[width=\columnwidth]{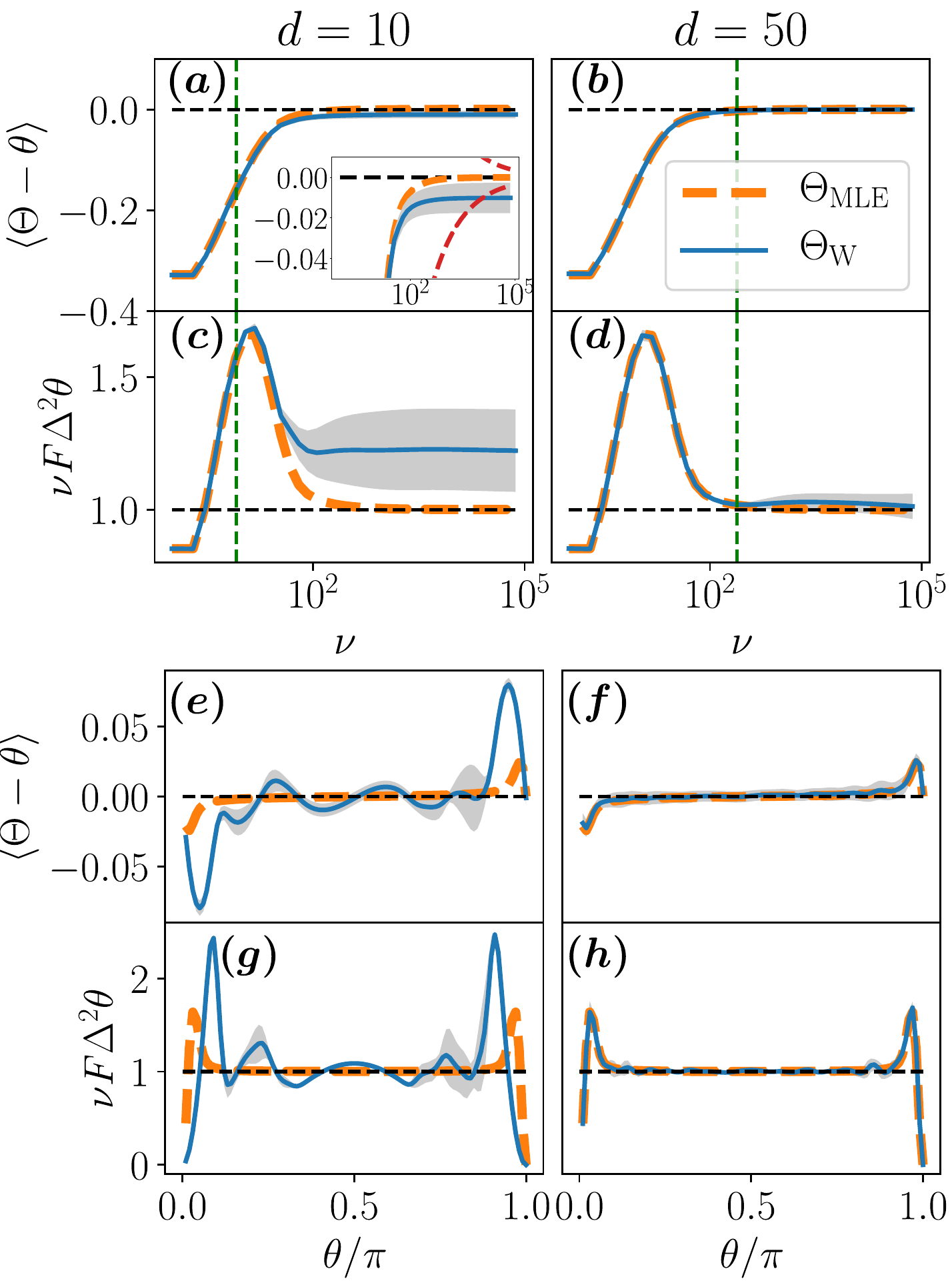}
\caption{\textbf{The grid-resolution limits generalisability.} For a single qubit we train 10 models with $m=10^3$ until the cost saturates the phase-averaged CRB $C\sim 1/m$ and show here the mean result (solid blue) and standard deviation (shaded gray regions). We test the models and compute the bias \textbf{(a,b,e,f)} and mean-square error normalised to the CRB \textbf{(c,d,g,h)}, both as a function of $\nu$ at $\theta=0.2\pi$ (top) and as a function of $\theta$ at $\nu=500$ (bottom). The $\theta$ values used to generate this plot are deliberately distinct from the training grid. To explore the role of the grid resolution, we compare a grid with $d=10$ grid points (left) to $d=50$ (right) both between $\theta \in \left[0, \pi \right]$ and sampled from a flat prior.  In \textbf{(a-d)} we indicate the grid-resolution limit $\delta \theta = L/(d-1)$ with a vertical dashed green line. The models consistently agree well with the MLE, up to a point determined by the grid resolution where the results have large fluctuations relative to the bias. In the inset  \textbf{(a)}, we show that for $m$ large enough, the model fails totally, as indicated by non-zero bias by margin not accounted for by the variance $\Delta^2 \theta=1/\nu$ (dashed red lines). All models have a single hidden layer of $n_h=1024$ neurons. The training set contains $M_j=2 \times 10^3$ feature vectors at each grid point (on average) and during training we use a mini-batch size of 256 with 75 training epochs, which ensures the cost saturates the phase-averaged CRB Eq.~(\ref{eq:CRBphaseav}).}
\label{fig:generalise_m}
\end{figure}

In Fig.~\ref{fig:generalise_m} again for the single-qubit system, we plot the mean bias and variance of the machine-learned estimator $\Theta_W$ both as a function of $\nu$ for fixed $\theta$ (Fig.~\ref{fig:generalise_m}, a-d) and as a function of $\theta$ for a fixed $\nu$, for two different grid resolutions $\delta \theta = L/(d-1)$.
Training is performed until the cost function gets as close as possible to the phase-averaged CRB (as discussed in Fig.~\ref{fig:costsCRB}, here $\mathrm{CRB}_{\theta \mathrm{av}}=1/{m}$) indicating $\Theta_W \approx \Theta_{\rm MLE}$ as well as possible given the training set. In Fig.~\ref{fig:generalise_m}, (a-d) the resolution limit $1/F\sqrt{\nu} = \delta \theta$ ($F(\theta)=1$ for a single qubit) is shown as a vertical dashed-green line - and is an excellent predictor of the point beyond which the training starts to produce inconsistent results.  Specifically, we train 10 models, and indicate the average with a solid blue line, the shaded gray regions are one standard deviation. Beyond the resolution limit, the machine-learned model does not converge to a consistent estimator, and does not reliably saturate the CRB. Far enough beyond the resolution limit the model breaks down totally. In the inset we show that the bias is eventually non-zero by a margin not accounted for by the variance $\Delta^2 \theta=1/\nu$ (dashed red lines), indicating that the model can no longer be trusted. In the bottom panels Fig.~\ref{fig:generalise_m}, (e-h) we verify that $\Theta_W$ reproduces the MLE across a range of $\theta$ values not present in the training set. As expected, the model performs better when trained on a finer grid-resolution. To conclude, we might say that so long as
\begin{equation} \label{eq:condition1}
    \nu \lessapprox \frac{1}{F(\theta) \delta \theta^2}
\end{equation}
that a well-trained model can reliably replicate the MLE, and importantly, inherit its desirable asymptotic properties i.e. that it is unbiased and saturates the CRB.

\subsection{The role of $m$ relative to the grid resolution.}

\begin{figure}[t!]
\centering
\includegraphics[width=\columnwidth]{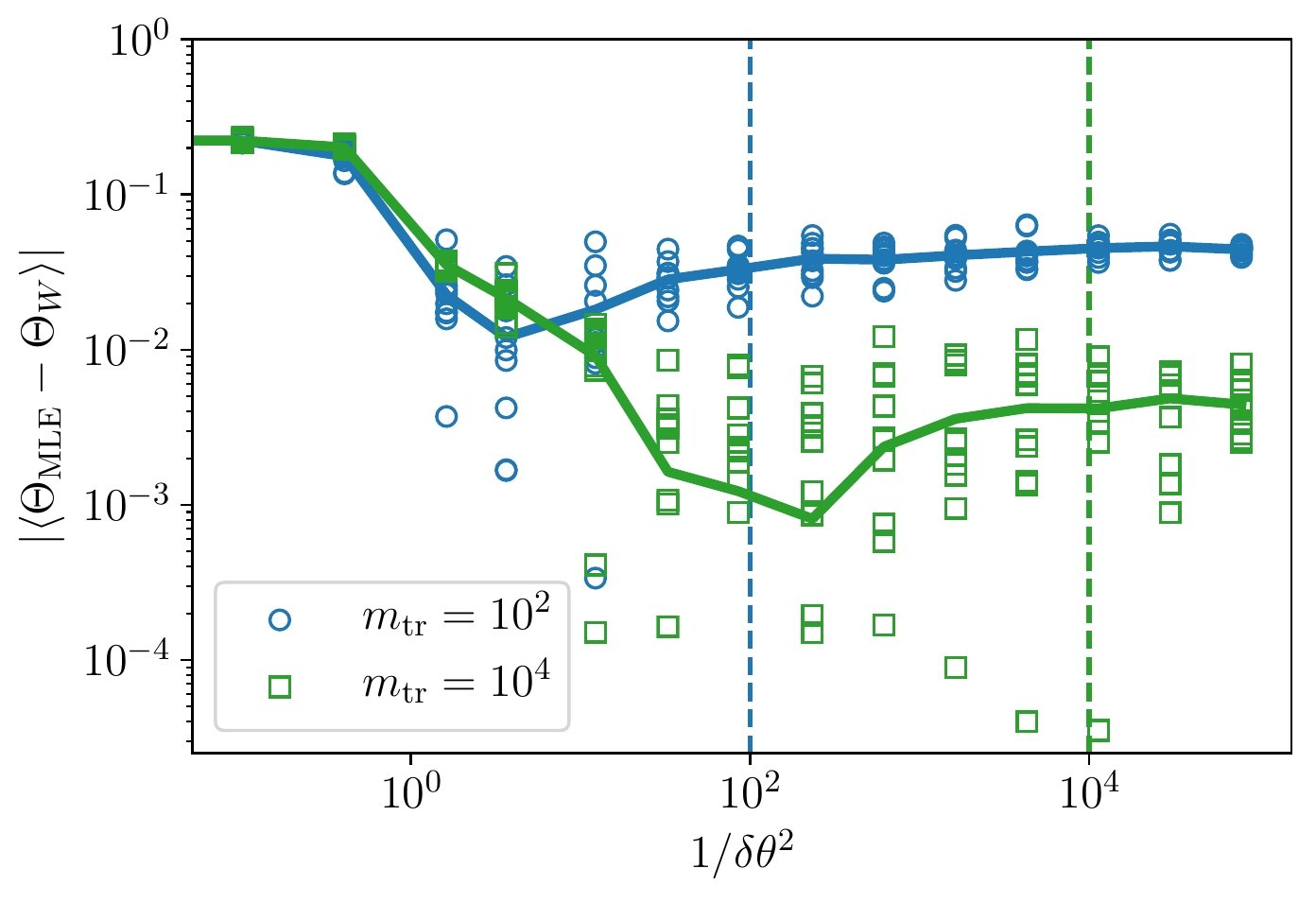}
\caption{\textbf{Quantum noise in the training data limits the grid resolution}. For $\nu=10$ shots at $\theta=0.2\pi$ we explore how close the machine learned estimator can come to the true MLE as a function of the number of grid-points $d$, for a single qubit. To account for variance in the randomly generated initial networks, we randomly generate 10 initial models, and then train each using a single training set for each $d$  (hollow points), the mean is the solid line. Beyond a certain $d$, a finer grid yields no additional advantage, and can even result in worse performance on average. The reason is quantum noise in the training data, each point has a width $\sim 1/\sqrt{F m}$, which can exceed the grid resolution $\delta \theta = L/(d-1)$. This noise limit is indicated by the dashed vertical lines. For each $d$, the training set and model architecture are identical to those used in Fig.~(\ref{fig:generalise_m}).}
\label{fig:varyd}
\end{figure}

Finally, we explore the role of quantum noise in the training data. The number of training measurements $m$ performed at each $\theta$ results in a variance bounded by $\Delta^2 \theta_{\rm CRB}(\theta) = 1/m F(\theta)$ when $m$ is large. This effectively sets a resolution limit for the training grid, beyond which neighbouring grid points cannot be distinguished. In Fig.~\ref{fig:varyd}, assuming the MLE is the expected result, we plot the distance between the machine-learned estimator $\Theta_W$ to the MLE. As the resolution increases, $\Theta_W$ comes closer to learning the MLE, but only up to a point. Beyond this point, there is no additional advantage in a finer training grid. We train 10 models for $m=10^2$ and $m=10^4$, to demonstrate that reducing the noise in the training set - which allows for a finer resolvable grid - produces an estimator that comes closer to the MLE. In other words, the number of training measurements sets a noise limit on the minimum useful grid resolution
\begin{equation} \label{eq:condition2}
    m \gtrapprox \frac{1}{F(\theta) \delta \theta^2},
\end{equation}
which in turn sets the maximum possible $\nu$ for which the model is reliably unbiased and saturates the CRB.


\section{Non-classical States of Many Qubits}

\begin{figure}[t!]
\centering
\includegraphics[width=\columnwidth]{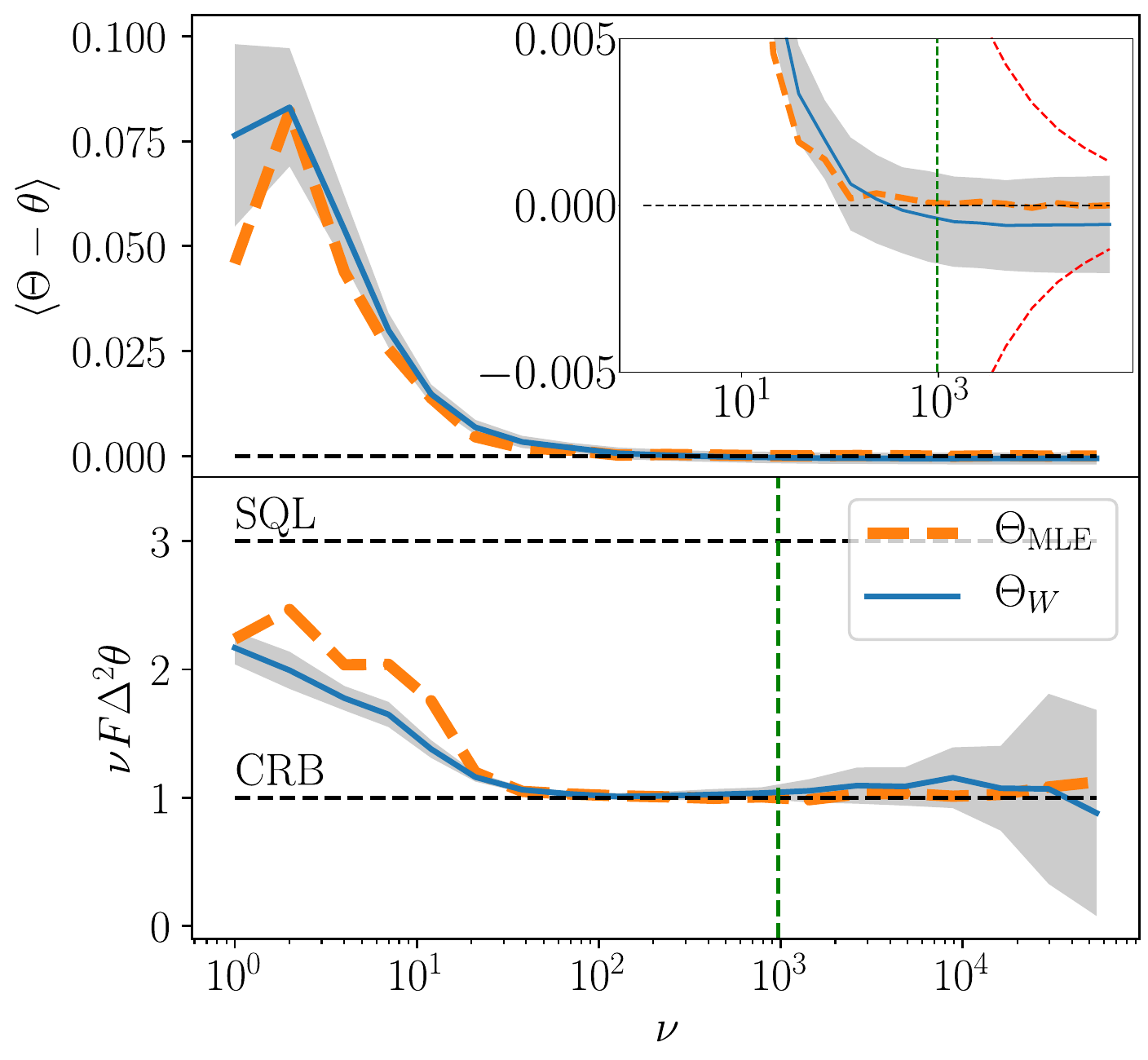}
\caption{\textbf{Non-classical $N$-qubit state.} We compute the bias (top) and variance (bottom) for an $N=4$ twin-Fock state (TFS), compared to the SQL and CRB (black dashed lines). We train 10 models and compare the mean (solid blue) to the MLE (dashed orange). Shaded gray regions represent the standard deviation in the models. Models are trained in the regime where the conditions Eq.~(\ref{eq:conditionprior}), Eq.~\ref{eq:condition1} and Eq.~\ref{eq:condition2} are all met, namely $m=10^3$ training measurements with $d=50$ grid-points on $[0,\pi/2]$. For training we use $M_j=10^3$ feature vectors per grid-point, and terminate training once the phase-averaged CRB is saturated, here roughly after 400 epochs with a mini-batch size of 128. The ANNs have $N+1=5$ input neurons, and 2 hidden layers with 128 neurons each. For the plot, we fix $\theta=0.2 \pi$ and approximate expectation values by averaging over $10^4$ randomly chosen measurement sequences $\boldsymbol{\mu}$ at each $m$. The grid-resolution $\delta \theta = \pi/2(d-1)$ is a green vertical dashed line. \textbf{(inset)} Zoom-in of the bias, showing that at large $m$ the fluctuations in the trained models are large relative to the root-variance $\Delta \theta$ for the TFS, computed exactly (dashed red curves).}
\label{fig:TFSMLE}
\end{figure}

Thus far we have made an extensive study of a single qubit, mainly to build understanding. However, it is important to study more complex states. In particular, there is currently a great effort to engineer many-qubit states with non-classical correlations, capable of providing sensitivities beyond the SQL \cite{PezzeRMP2018}. The limitations explored in the previous section must be revisited with these more complex states in mind. Specifically, the criteria Eq.~\ref{eq:condition1} is not a strict cut-off. Rather, it should be interpreted as a regime where the model can no longer reliably reproduce the MLE. Eventually, this manifests as large fluctuations in the point estimates $\Theta_W$ provided by different models with identical training. Combining the regimes Eq.~(\ref{eq:condition1}) and Eq.~(\ref{eq:condition1}) one can deduce that $\nu < m$, which limits the usefulness of the model. If we must collect many more training measurements $m$ than we could ever actually use in a real application, the model would be wasteful in some sense. That the model performs well (i.e. it has sub SQL variance and is unbiased) for some $\nu$ beyond the resolution limit is important for this reason. In this section we verify that machine-learned estimators can yield unbiased sub-shot noise sensitivities for complex non-classical states even beyond the resolution limit - effectively demonstrating that these models can usefully generalise from the training set.

In Fig.~\ref{fig:TFSMLE} we compute the variance and bias for a twin-Fock state (TFS) as a function of the number of measurements $m$ at a fixed phase. An $N$-qubit TFS is defined by a symmetrized combination of $N/2$ spin-up and $N/2$ spin-down particles $|\rm TFS \rangle = {\rm Symm}\{ \ket{\downarrow}^{\otimes N/2}, \ket{\downarrow}^{\otimes N/2} \}$. We consider a rotation of this state by the unitary $U=\exp( -i \hat{J}_y \theta)$, where the collective spin operator is defined $\hat{J}_k=\sum_{i=1}^N \sigma_k^{(i)}/2$, where $\sigma_k^{(i)}$ is the $k$th Pauli matrix for the $i$th qubit. The operation $\hat{U}$ is equivalent to a Mach-Zehnder interferometer~\cite{PezzeRMP2018}. The CRB for this system is $\Delta^2 \theta_{\rm CRB} = 1/[\nu N/2(N/2+1)]$ which is below the SQL $\Delta^2 \theta_{\rm SQL} = 1/\nu N$. In Fig.~\ref{fig:TFSMLE} we train 10 models $\Theta_W$ until the phase-averaged CRB is saturated, and compare the mean (solid blue) and standard deviation (shaded gray region) to the MLE (dashed orange). Above the resolution of the training grid, the models exhibit large variance, similarly to the single-qubit example in Fig.~\ref{fig:generalise_m}. However, the variance remains comfortably below the SQL and the bias remains zero within the variance $\Delta^2 \theta$ (computed exactly for the TFS and plotted in dashed red, see Fig.~\ref{fig:generalise_m} inset). This important benchmark indicates that useful sub-SQL parameter estimation is possible using an ANN estimator, and that the estimator generalises well beyond the $\nu$ used for training ($m=10^3$ in this case). Similar results were reported in Ref. \cite{CiminiPRL2019}, but not at the CRB, presumably due to the resolution of the training grid.

One final observation is that for the TFS, unlike the single-qubit example, the model generalises poorly to $\nu \ll m$. See for instance in Fig.~\ref{fig:generalise_m} that between $\nu \sim 1-10$ (note $m=10^3$) the machine-learned estimators differ significantly from the MLE. The reason is simply that the TFS likelihood function is extremely non-Gaussian for small $\nu$, but was trained in the large $\nu$ limit (such that Eq.~\ref{eq:post_asymptotic} applies). As a result, feature vectors $\boldsymbol{\mu}$ with relatively few measurements differ significantly from the training examples and are therefore difficult for the model to correctly label. This is not a major concern, as experiments are typically not interested in the low $\nu$ regime which is generally highly biased with sensitivity far from the CRB (see the dashed-orange curve). 

\section{Conclusion}

In this manuscript we have explored the training of a machine-learned point estimate using artificial neural networks. Using a simple single-qubit example, we show that the prior knowledge plays an important role in determining which estimator the machine will ``decide'' to learn - especially when the number of training measurements are small. So long as the Bayesian posterior distribution over the training set is single-peaked, smooth and non-zero around $\theta$, this estimator is a good approximation to the Bayesian maximum a-posteriori estimator. We regard this result - that prior knowledge is an inescapable part of the training process - as evidence that supervised learning itself is naturally Bayesian.

When the number of training measurements are large, this Bayesian estimator coincides with the well-known maximum-likelihood estimator. 
Based on this, we argue that during training the cost function should converge to the well-known Cram{\'e}r–Rao bound. 
If this bound is known a-priori, it can be used to select optimal training parameters (such as the number of hidden neurons in the network, the batch-size, number of training epochs, etc.). 
We then argue that so long as the Fisher information of the prior is large relative to the CRB, the resolution of the training grid is key to building a model that generalises well to large $m$. Furthermore, we show that quantum noise in the training data imposes a fundamental limit to the minimum useful resolution of the training grid. Finally, using a specific example we show that ANN estimators can provide unbiased estimates with sensitives below the classical limit up to a number of measurements larger than those present in the training set, indicating that the model can generalise well. Our results pave the way for maximum-likelihood inference to play a more significant role in the operation of quantum sensors. Currently, MLE is relegated merely to a  handful of proof-of-principle experiments
\cite{KrischekPRL2011, XiangNATPHOT2011, SinclairPRA2017, XuPRL2020} due to challenges associated with device calibration.

\section{Acknowledgments}

We would like to thank V. Gebhart for useful discussions. We acknowledge funding from the project EMPIR-USOQS, EMPIR projects are co-funded by the European Unions Horizon2020 research and innovation programme and the EMPIR Participating States. 
We also acknowledge financial support from the European Union's Horizon 2020 research and innovation programme - Qombs Project, FET Flagship on Quantum Technologies grant no. 820419, and from the H2020 QuantERA ERA-NET Cofund in Quantum Technologies projects QCLOCKS and CEBBEC.


\begin{thebibliography}{100}
\bibitem{note_class}
Strictly speaking, we are talking about probabilistic classification - the model is trained to provide a conditional probability that a possible category is correct given the input data. Non-probabilistic classifiers do exist (e.g. support-vector machines). Like a regression model, these would also need to be analysed in a frequentist way. The model would provide only a point estimate (in this case a discrete category) and not a confidence interval. However, the parameter to be estimated is typically continuous (e.g a phase) and so we restrict ourselves to regression in this manuscript.
\bibitem{MSE}
The mean-square error $\Delta^2 \theta_{\rm MSE} = \langle (\Theta(\boldsymbol{\mu}) - \theta)^2\rangle$ of an estimator $\Theta$ is related to the variance $\Delta^2 \theta = \langle \Theta(\boldsymbol{\mu}) - \langle \Theta \rangle )^2\rangle$ by $\Delta^2 \theta_{\rm MSE} = \Delta^2 \theta + (\langle \Theta \rangle-\theta)^2$. The bias term is strictly positive so any bound on the variance is also a bound on the mean-square error.
\end{thebibliography}
\end{document}